\documentclass[preprintnumbers,article,amsmath,amssymb,floatfix,10pt,prd,twocolumn,
superscriptaddress,nofootinbib]{revtex4-2}
\usepackage{bm}
\usepackage{amsfonts}
\usepackage{latexsym}
\usepackage[latin1]{inputenc}
\usepackage{graphicx}
\usepackage{amsmath}
\usepackage{palatino}
\usepackage{mathpazo}
\usepackage{textcomp}
\usepackage{float}
\usepackage{booktabs}
\usepackage{dcolumn}
\usepackage{ragged2e}
\usepackage{hyperref}
\hypersetup{colorlinks,citecolor=blue}
\hypersetup{colorlinks=true,linkcolor=red,filecolor=magenta,    urlcolor=blue}
\usepackage{amsmath}
\usepackage{xcolor}
\usepackage{orcidlink}
\usepackage{epsfig}
\usepackage{subfigure}
\usepackage{commath}

\def\jnl@style{\it}
\def\aaref@jnl#1{{\jnl@style#1}}

\def\aaref@jnl#1{{\jnl@style#1}}

\def\aj{\aaref@jnl{AJ}}                   
\def\apj{\aaref@jnl{ApJ}}                 
\def\apjl{\aaref@jnl{ApJ}}                
\def\apjs{\aaref@jnl{ApJS}}               
\def\apss{\aaref@jnl{Ap\&SS}}             
\def\aap{\aaref@jnl{A\&A}}                
\def\aapr{\aaref@jnl{A\&A~Rev.}}          
\def\aaps{\aaref@jnl{A\&AS}}              
\def\mnras{\aaref@jnl{Mon.~Not.~Roy.~Astron.~Soc.}}             
\def\prd{\aaref@jnl{Phys.~Rev.~D}}        
\def\prc{\aaref@jnl{Phys.~Rev.~C}}  
\def\prl{\aaref@jnl{Phys.~Rev.~Lett.}}    
\def\qjras{\aaref@jnl{QJRAS}}             
\def\skytel{\aaref@jnl{S\&T}}             
\def\ssr{\aaref@jnl{Space~Sci.~Rev.}}     
\def\zap{\aaref@jnl{ZAp}}                 
\def\nat{\aaref@jnl{Nature}}              
\def\aplett{\aaref@jnl{Astrophys.~Lett.}} 
\def\apspr{\aaref@jnl{Astrophys.~Space~Phys.~Res.}} 
\def\physrep{\aaref@jnl{Phys.~Rep.}}      
\def\physscr{\aaref@jnl{Phys.~Scr}}       
\def\commat{\aaref@jnl{Comm.~Math.~Phys.}}              
\def\science{\aaref@jnl{Science}}               
\def\cqg{\aaref@jnl{Classical Quant.~Grav.}}            
\def\jpcs{\aaref@jnl{JPCS}}                                     
\def\ijmpd{\aaref@jnl{Int.~J.~Mod.~Phys.~D}}                    
\def\grg{\aaref@jnl{Gen.~Relat.~Gravit.}}               
\def\rpp{\aaref@jnl{Rep.~Prog.~Phys.}}          
\def\npa{\aaref@jnl{Nucl.~Phys.~A}}        
\def\lrr{\aaref@jnl{Living Rev.~Rel.}}                   
\def\jcap{\aaref@jnl{J.~Cosmology Astropart.~Phys.}}    
\def\rmp{\aaref@jnl{Rev.~Mod.~Phys.}}   
\def\epjc{\aaref@jnl{Eur.~Phys.~J.~C}} 
\def\plb{\aaref@jnl{~Phy.~Lett.~B}} 
\def\mpla{\aaref@jnl{Mod.~Phy.~Lett.~A}} 
\def\arxiv{\aaref@jnl{arxiv.org}}

\allowdisplaybreaks[1]

\begin{document}
\color{black}       
\title{Deflection of light by wormholes and its shadow due to dark matter within modified symmetric teleparallel gravity formalism}

\author{G. Mustafa\orcidlink{0000-0003-1409-2009}}
\email{gmustafa3828@gmail.com}
\affiliation{Department of Physics, Zhejiang Normal University, Jinhua, 321004, People's Republic of China.}

\author{Zinnat Hassan\orcidlink{0000-0002-6608-2075}}
\email{zinnathassan980@gmail.com}
\affiliation{Department of Mathematics, Birla Institute of Technology and
Science-Pilani,\\ Hyderabad Campus, Hyderabad-500078, India.}

\author{P.K. Sahoo\orcidlink{0000-0003-2130-8832}}
\email{pksahoo@hyderabad.bits-pilani.ac.in}
\affiliation{Department of Mathematics, Birla Institute of Technology and
Science-Pilani,\\ Hyderabad Campus, Hyderabad-500078, India.}

%
\date{\today}
\begin{abstract}
We explore the possibility of traversable wormhole formation in the dark matter halos in the context of $f(Q)$ gravity. We obtain the exact wormhole solutions with anisotropic matter source based on the Bose-Einstein condensate, Navarro-Frenk-White, and pseudo-isothermal matter density profiles. Notably, we present a novel wormhole solution supported by these dark matters using the expressions for the density profile and rotational velocity along with the modified field equations to calculate the redshift and shape functions of the wormholes. With a particular set of parameters, we demonstrate that our proposed wormhole solutions fulfill the flare-out condition against an asymptotic background. Additionally, we examine the energy conditions, focusing on the null energy conditions at the wormhole's throat, providing a graphical representation of the feasible and negative regions. Our study also examines the wormhole's shadow in the presence of various dark matter models, revealing that higher central densities result in a shadow closer to the throat, whereas lower values have the opposite effect. Moreover, we explore the deflection of light when it encounters these wormholes, particularly noting that light deflection approaches infinity at the throat, where the gravitational field is extremely strong.
\end{abstract}

\maketitle


\section{Introduction}\label{sec1}
\indent The theory of general relativity and other extended theories of gravity present the possibility of space-time hosting complicated structures, such as wormholes. These wormholes are tunnel-like passages that connect disparate or distant regions of space. Within the General Relativity (GR) framework, black holes and wormholes stand out as fascinating astrophysical phenomena. While researchers have confirmed the existence of black holes in \cite{Abbott1,Abbott2,Abbott3}, the presence of wormholes remains an ongoing investigation. The concept of wormholes traces back to the pioneering work of Einstein and Rosen, who proposed the first wormhole solution known as the Einstein-Rosen bridge \cite{Rosen1}. Wormholes gained renewed interest when Ellis \cite{Rosen2} introduced a novel solution incorporating a spherically symmetric configuration of Einstein's equations, incorporating a massless scalar field with ghost properties. Morris and Thorne \cite{Thorne/1988} later demonstrated that these Ellis wormholes could indeed be traversable, potentially facilitating rapid travel through space and even raising the possibility of time travel. Notably, such wormhole models do not feature singularities or horizons, and their tidal forces are deemed survivable for humans. Additionally, Morris and Thorne \cite{Thorne/1988} indicated that these wormhole solutions would violate the null energy conditions, necessitating the presence of exotic matter. This exotic matter, disobeying energy conditions, exhibits characteristics that challenge established laws of physics, including the potential for particles to possess negative mass. An extensive investigation has been conducted into the presence of wormholes in \cite{Khatsymovsky1}, while numerous scholars have explored the stability of traversable wormholes. Notably, Shinkai and Hayward \cite{Khatsymovsky2} demonstrated the instability exhibited by Ellis wormholes through numerical simulations. Since the inception of the traversable wormhole concept, researchers have been fascinated by the possibility of constructing such passages using ordinary matter. Recent research indicates that within modified gravity theories, it may be feasible to construct wormholes composed of ordinary matter that satisfy all energy conditions \cite{Khatsymovsky3}. However, in the process of utilizing modified gravity to create wormholes, although the matter involved may be ordinary, the effective geometric matter, which serves as the source of modified gravity, could still violate the standard null energy condition. Several investigations have been done on wormhole configurations that do not demand exotic matter \cite{Tanaka1,Tanaka2,Tanaka3,Tanaka4,Tanaka5,Tanaka6,Tanaka7,Tanaka8}.\\
\indent The formulation of $f(Q)$ gravity arises from the development of novel classes of modified gravity, originating from symmetric teleparallel gravity based on the non-metricity scalar Q. Essentially, $f(Q)$ gravity serves as an extension of symmetric teleparallel gravity, which operates within a framework of a flat connection with vanishing torsion. Jimenez et al. \cite{Jimenez} introduced a theoretical framework where curvature and torsion vanish, and gravity is attributed to non-metricity $Q$. This $f(Q)$ theory has demonstrated the capacity to explain the Universe's accelerated expansion with a statistical precision comparable to renowned modified gravities \cite{Lin1}. The cosmological implications of $f(Q)$ gravity have been extensively explored in \cite{Lin2,Lin4,Lin5}. Harko and his collaborators \cite{Lin3} employed $f(Q)$ gravity to study cosmological evolutions and related phenomena. Additionally, Anagnostopoulos et al. \cite{Anagnostopoulos} employed Big Bang Nucleosynthesis formalism and observational data to constrain various classes of $f(Q)$ models.\
Investigations into black hole solutions within the framework of $f(Q)$ gravity have been conducted in Refs. \cite{Shaun1,Shaun2,Shaun3}. Additionally, the formation and properties of compact stars resulting from gravitational decoupling in $f(Q)$ gravity theory have been examined in \cite{Shaun4}. Sokoliuk et al. \cite{Shaun5} studied Buchdahl quark stars within the context of the $f(Q)$ theory. Spherically symmetric configurations within $f(Q)$ gravity have been investigated in \cite{Lin1}. They have examined with a specific polynomial expression, such as $f(Q) =Q+\alpha Q^2$, while employing polytropic Equations of State (EoS) to characterize internal spherically symmetric configurations. Wang et al. \cite{Wang1} presented static and spherically symmetric solutions incorporating an anisotropic fluid for general $f(Q)$ gravity formulations. Calza and Sebastiani \cite{Wang2} analyzed a class of topological static spherically symmetric vacuum solutions with constant non-metricity within $f(Q)$ gravity.
\indent Further, Mustafa et al. \cite{Wang3} employed the Karmarkar conditions in the $f(Q)$ gravity formalism to derive wormhole solutions that adhere to Energy Conditions (ECs). Furthermore, investigations into wormhole geometries within the $f(Q)$ gravity framework have been carried out in \cite{Wang4}, revealing that a linear model may require a minimal amount of exotic matter for a traversable wormhole. Additionally, Sharma et al. \cite{Wang5} explored wormhole solutions within symmetric teleparallel gravity, emphasizing specific shape and redshift functions within certain $f(Q)$ models that can yield solutions satisfying ECs in some regions of space-time. Recently, the Casimir wormhole and its GUP correction in the $f(Q)$ gravity framework have been examined in \cite{Ghosh1,Ghosh2}. Moreover, readers can also check some interesting works of literature related to astrophysical objects found in non-metricity-based modified theories of gravity (see Refs. \cite{Pradhan1,Pradhan2,Pradhan3,Pradhan4,Pradhan5,Pradhan6}).\\
\indent The notion of dark matter, a mysterious form of matter containing approximately 25\% of the Universe's total matter content, emerges from observational predictions. Various candidates from particle physics and supersymmetric string theory, such as axions and weakly interacting massive particles, are considered compelling nominees for dark matter despite the absence of direct experimental confirmation. Nevertheless, indications of its presence are observed in phenomena such as galactic rotation curves \cite{Rubin1}, galaxy cluster dynamics \cite{Rubin2}, and cosmological observations of anisotropies in the cosmic microwave background as measured by PLANCK \cite{Rubin3}. The literature \cite{Rahaman1,Rahaman2} explores considerations of traversable wormholes (TWs) within dark matter halos and galaxy formation regions, typically based on the Navarro-Frenk-White (NFW) profile \cite{Rahaman3} of matter distribution. Rahaman et al. \cite{Rahaman4} initially proposed the existence of potential wormholes in the outer regions of galactic halos based on the NFW density profile, extending their analysis to utilize the universal rotation curve (URC) dark matter model to derive analogous results within the central portion of the halo \cite{Rahaman2}. Also, dark matter, considered a non-relativistic matter describable by NFW and King profiles, is employed to construct wormholes \cite{Rahaman6}. Discrepancies between NFW halo velocity profiles and the observed dynamics of spiral galaxies remain unresolved, leading to modifications in the original NFW halo profiles within the $\Lambda$CDM scenario to align with observational data \cite{Rahaman5}. Additionally, it is shown that the presence of TWs in nature could be inferred through the study of scalar wave scattering \cite{Rahaman1}.\\
\indent In a recent study, Jusufi et al. \cite{Jusufi1} highlighted the potential formation of TW through the presence of a Bose-Einstein condensation dark matter (BEC-DM) halo. This BEC-DM model presents a more reasonable framework, particularly concerning the smaller scales of galaxies when compared to the Cold Dark Matter (CDM) model \cite{Jusufi2}. Notably, within the inner regions of galaxies, the interactions among dark matter particles are significantly vital, resulting in a deviation from cold dark matter behavior and rendering the density profile unsuitable. Consequently, the BEC-DM model indicates considerably lower dark matter densities in the central regions of galaxies compared to those projected by the NFW profile. Additionally, an alternative category of dark matter characterized by a pseudo isothermal (PI) profile, alongside the CDM and BEC-DM model, is associated with modifications to gravity, such as Modified Newtonian Dynamics (MOND) \cite{Jusufi3}. MOND \cite{Milgrom1,Milgrom2,Milgrom3} indicates that the discrepancies in mass within galactic systems arise not from dark matter but from deviations from standard dynamics at lower accelerations. In \cite{Milgrom4}, Paul investigated the existence of TWs in the presence of MOND with or without a scalar field.\\
The exploration of shadows cast by compact objects in astrophysics has become a pivotal research focus, providing valuable insights into the intrinsic properties of these objects and the fabric of spacetime \cite{sd1,sd2,sd3,sd4}. Observations of the shadows of compact objects, particularly those at the core of $SgrA^*$ and $M87$, have opened new routes for testing theories of gravity and examining various astrophysical models \cite{Abbott3,sd6}. Among these, the shadow of a wormhole is particularly fascinating, offering a theoretical means to explore the universe's structure in novel ways. Notably, the shadow of a wormhole could offer indirect evidence of these hypothetical spacetime tunnels, distinguishing itself from other compact objects. Analyzing the wormhole's shadow could yield valuable data on its structure, including throat size, spin, and potential accretion processes. The analysis of shadows from various objects, including black holes and wormholes, across different gravity models, with and without the presence of plasma, are extensively covered in \cite{sd7,sd8,sd9,sd10,sd11,sd12,sd13,sd4,sd15}.\\
\indent Gravitational lensing stands as an early useful exploration within the realm of general relativity, which was initially delved into by Einstein \cite{Einstein10}. This phenomenon unfolds when a significantly massive celestial body bends incoming light, much like a lens, offering observers enhanced insights into the originating source. The interest in this area surged following the observed validation of light's deflection, as indicated theoretically \cite{Dyson,Eddington}. Its scope extends beyond celestial bodies, opening avenues to probe into exoplanets, dark matter, and dark energy. A notable milestone was achieved with the first successful measurement of a white dwarf's mass, Stein 2051 B, through astrometric microlensing \cite{Sahu}. An intriguing aspect of gravitational lensing is its potential to cause light to bend infinitely in unstable light rings, creating numerous relativistic images under strong lensing conditions \cite{Bozza1,Virbhadra,Bozza22}. This phenomenon, in both strong and weak forms, serves as a potent analytical tool for examining gravitational fields near various cosmic entities, including black holes and wormholes. Theoretical and astrophysical investigations have extensively applied gravitational lensing to study wormholes, reflecting its significance in contemporary research \cite{A1,A2,A3,A4,A5,A6,A7,A8,A9,A10}.\\
\indent Motivated by the above discussions, we explore wormhole solutions under different DM halo models within $f(Q)$ gravity. The paper is structured as follows: Section \ref{sec2} covers the basic formalism of $f(Q)$ gravity and the associated wormhole field equations. Section \ref{sec3} discusses dark matter profiles and wormhole solutions, while Section \ref{sec4} presents the energy conditions analytically and graphically. Wormhole shadows and deflection angles are examined in Sections \ref{sec5} and \ref{sec6}, respectively. Section \ref{sec7} details the geometry of embedded wormhole spacetime with diagrams. The paper concludes with final discussions.
\section{Wormhole Field equations in $f(Q)$ gravity}
\label{sec2}
In this section, we aim to introduce the fundamental layout of the wormhole theory and provide a brief review of the $f(Q)$ gravity formalism. We consider spherically symmetric static Morris-Thorne wormhole metric  \cite{Thorne/1988}, defined by
\begin{equation}\label{3a}
ds^2=-e^{2\phi(r)}dt^2+\left(1-\frac{b(r)}{r}\right)^{-1}dr^2+r^2d\theta^2+r^2\text{sin}^2\theta d\Phi^2,
\end{equation}
where $b(r)$ and $\phi(r)$ represent the shape and redshift functions, respectively. The flaring-out condition is a key factor in determining whether a wormhole can be traversed. This condition is mathematically expressed as \cite{Thorne/1988}
\begin{equation}
(b-b'r)/b^2>0.
\end{equation}
Specifically, at the wormhole's throat, this formula simplifies the requirement that 
\begin{equation}
 b'(r_0)<1.
\end{equation}
Additionally, for any point beyond the throat, i.e., $r>r_0$, the condition $1-\frac{b(r)}{r}>0$ must also be met. Further, the wormhole should be asymptotically flat, i.e., 
\begin{equation}
\frac{b(r)}{r}\rightarrow 0 \,\,\,\, \text{as}\,\,\,r\rightarrow 0.
\end{equation}
Also, for an event horizon-free, the redshift function must be finite everywhere.

Now, we will provide an overview of $f(Q)$ gravity, also known as symmetric teleparallel gravity, as initially proposed by Jimenez et al. \cite{Jimenez}. The formulation of this gravity involves an action expressed as
\begin{equation}\label{6a}
\mathcal{S}=\int\frac{1}{2}\,f(Q)\sqrt{-g}\,d^4x+\int \mathcal{L}_m\,\sqrt{-g}\,d^4x\, ,
\end{equation}
where $f(Q)$ represents an arbitrary function of $Q$ and $\mathcal{L}_m$ denotes the matter Lagrangian density. $g$ corresponds to the determinant of the metric tensor $g_{\mu\nu}$.\\
Now, one can define the non-metricity tensor\\
\begin{equation}\label{6b}
Q_{\lambda\mu\nu}=\bigtriangledown_{\lambda} g_{\mu\nu}=\partial_\lambda g_{\mu\nu}-\Gamma^\beta_{\,\,\,\lambda \mu}g_{\beta \nu}-\Gamma^\beta_{\,\,\,\lambda \nu}g_{\mu \beta},
\end{equation}
where, $\Gamma^\beta_{\,\,\,\mu\nu}$ is the metric affine connection.\\
Also, the superpotential is expressed by
\begin{multline}\label{6c}
P^\alpha\;_{\mu\nu}=\frac{1}{4}\left[-Q^\alpha\;_{\mu\nu}+2Q_{(\mu}\;^\alpha\;_{\nu)}+Q^\alpha g_{\mu\nu}-\tilde{Q}^\alpha g_{\mu\nu}\right.\\\left.
-\delta^\alpha_{(\mu}Q_{\nu)}\right],
\end{multline}
where
\begin{equation}
\label{6d}
Q_{\alpha}=Q_{\alpha}\;^{\mu}\;_{\mu},\; \tilde{Q}_\alpha=Q^\mu\;_{\alpha\mu}.
\end{equation}
are two traces.\\
Further, the non-metricity scalar is presented by \cite{Jimenez}
\begin{equation}
\label{6e}
Q=-Q_{\alpha\mu\nu}\,P^{\alpha\mu\nu}.
\end{equation}
In this context, the field equations are derived by varying the action \eqref{6a} with respect to the metric tensor $g_{\mu\nu}$
\begin{multline}\label{6f}
\frac{2}{\sqrt{-g}}\bigtriangledown_\gamma\left(\sqrt{-g}\,f_Q\,P^\gamma\;_{\mu\nu}\right)+\frac{1}{2}g_{\mu\nu}f \\
+f_Q\left(P_{\mu\gamma i}\,Q_\nu\;^{\gamma i}-2\,Q_{\gamma i \mu}\,P^{\gamma i}\;_\nu\right)=-T_{\mu\nu},
\end{multline}
where $f_Q=\frac{df}{dQ}$ and $T_{\mu\,\nu}$ is the energy-momentum tensor of the form
\begin{equation}\label{6g}
T_{\mu\nu}=-\frac{2}{\sqrt{-g}}\frac{\delta\left(\sqrt{-g}\,\mathcal{L}_m\right)}{\delta g^{\mu\nu}}.
\end{equation}
Further, one can vary the action with respect to the connection and obtain the extra constraint
\begin{equation}\label{6h}
\bigtriangledown_\mu \bigtriangledown_\nu \left(\sqrt{-g}\,f_Q\,P^\gamma\;_{\mu\nu}\right)=0.
\end{equation}
One can study this theory using a coincident gauge involving a specific coordinate choice. In this gauge, the connection disappears, and the non-metricity expressed in Eq. \eqref{6b} can be simplified to the form
\begin{equation}\label{2222}
Q_{\lambda\mu\nu}=\partial_\lambda g_{\mu\nu}.
\end{equation}
This simplification makes calculations easier since only the metric is considered a fundamental variable. However, it should be noted that the action is no longer diffeomorphism invariant in this case, except for standard GR, as stated in \cite{Lin2}.\\
Now, we can obtain the non-metricity scalar for the metric \eqref{3a} from Eq. \eqref{6e}
\begin{equation}
\label{12}
Q=-\frac{2}{r}\left(1-\frac{b}{r}\right)\left(2\phi^{'}+\frac{1}{r}\right).
\end{equation}
Also, we assume the matter that is described by the anisotropic fluid which can be written in the form
\begin{equation}\label{3b}
T^{\mu}_{\nu}=\text{diag}[-\rho,\,P_r,\,P_t,\,P_t],
\end{equation}
where $\rho$ represents the energy density. $P_r$ and $P_t$ denotes the radial and tangential pressure, respectively.\\
Now, the field equations for the metric \eqref{3a} under anisotropic matter \eqref{3b} within the framework of modified symmetric teleparallel gravity can be derived as
\begin{multline}
\label{14}
\left[\frac{1}{r}\left(-\frac{1}{r}+\frac{rb^{'}+b}{r^2}-2\phi^{'}\left(1-\frac{b}{r}\right)\right)\right]f_Q \\
-\frac{2}{r}\left(1-\frac{b}{r}\right)f_{QQ}Q^{'}-\frac{f}{2}=-\rho,
\end{multline}
\begin{equation}
\label{15}
\left[\frac{2}{r}\left(1-\frac{b}{r}\right)\left(2\phi^{'}+\frac{1}{r}\right)-\frac{1}{r^2}\right]f_Q+\frac{f}{2}=-P_r,
\end{equation}
\begin{multline}
\label{16}
\left[\frac{1}{r}\left(\left(1-\frac{b}{r}\right)\left(\frac{1}{r}+\phi^{'}\left(3+r\phi^{'}\right)+r\phi^{''}\right)-\frac{rb^{'}-b}{2r^2}\right.\right.\\\left.\left.
\left(1+r\phi^{'}\right)\right)\right]f_Q+\frac{1}{r}\left(1-\frac{b}{r}\right)\left(1+r\phi^{'}\right)f_{QQ}Q^{'}\\
+\frac{f}{2}=-P_t,
\end{multline}
\begin{equation}\label{111}
\frac{\cot{\theta}}{2}f_{QQ}Q^{'}=0,
\end{equation}
where ${'}$ represents $\frac{d}{dr}$. The nonzero off-diagonal metric components obtained by the specific gauge choice for the field equations in the setting of $f(T)$ theory of gravity \cite{dz1} impose some constraints on the functional form of $f(T)$. As a consequence, one can put the same constraints on the functional form of the $f(Q)$ theory of gravity. In this regard, within the scope of anisotropic matter distribution, Wang and his co-authors \cite{Wang1} developed the potential functional forms for $f(Q)$ gravity in the framework of the static and spherically symmetric spacetime. Interestingly, they have proved that the exact form of Schwarzschild solution can be derived only when $f_{QQ}(Q)=0$, while the other related solutions obtained by taking nonmetricity term $Q^{\prime}=0$ or $Q=Q_0$, where $Q_0$ is constant, provides the deviation from the exact Schwarzschild solution. To solve the system of field equations for $f(Q)$ gravity theory and obtain self-gravitating solutions, we derive the functional form of $f(Q)$ by setting $f_{QQ}$ to zero as:
\begin{eqnarray}
f_{QQ}(Q)=0~\Rightarrow~f_{Q}(Q)=c_0~\Rightarrow~f(Q)=c_0+c_1 Q,~~~~  \label{eq26}
\end{eqnarray}
where $c_0$ and $c_1$ are constants. We would like to emphasize that, at this point, the compatibility of a static spherically symmetric spacetime with the coincident gauge can be achieved if one assumes the affine connection to be zero and $f(Q)$ gravity theory has vacuum solutions (i.e., $T_{\mu\nu}=0$). In this scenario, the off-diagonal component can be expressed as 
\begin{eqnarray}
\frac{\text{cot}\,\theta}{2}\,Q^\prime\,f_{QQ}=0, \label{eq27}
\end{eqnarray}
where $Q$ has been already provided in Eq. (\ref{6e}). As a result of Eq. \ref{eq26}, it is evident that $f(Q)$ must be linear, leading to the correct transformation of the equations of motion to $f_{QQ}=0$ \cite{dz3}. Therefore, in order to achieve a more generalized form of the spherically symmetric metric within a fixed coincident gauge, it is necessary to ensure compatibility with an affine connection $\Gamma^\beta_{\,\,\,\mu\nu}=0$ \cite{dz3}. Hence, in this study, we have chosen a linear functional form by setting $f_{QQ}=0$ to derive the equations of motion, making the spherically symmetric coordinate system compatible with the considered affine connection. Therefore, we get the revised field equations as follows:\\
\begin{equation}
\label{2b9}
\rho=\frac{c_0 }{2}-\frac{c_1  b'}{r^2},
\end{equation}
\begin{equation}
\label{2b10}
P_r=\frac{c_1  \left(2 r (b-r) \phi'+b\right)}{r^3}-\frac{c_0 }{2},
\end{equation}
\begin{multline}
\label{2b11}
P_t=\frac{c_1  \left(r \phi '+1\right) \left(r b'+2 r (b-r) \phi '-b\right)}{2 r^3}\\
+\frac{c_1  (b-r) \phi ''}{r}-\frac{c_0 }{2}.
\end{multline}
Now, in the following sections, we shall study wormhole geometry under the effect of different dark matter models.
\section{Wormhole solutions due to dark matters}\label{sec3}
In this section, we shall try to find the wormhole shape function as well as the redshift function and discuss the necessary properties of a traversable wormhole under the effect of dark matter. For our study, we will consider three well-known dark matter models, Bose-Einstein condensate, Psudo Isothermal, and Navarro-Frenk-White.

\subsection{Bose-Einstein condensate}
This section presents the theory of Bose-Einstein Condensate (BEC) dark matter as outlined in \cite{Harko}. It is worth noting that in a quantum system containing N interacting condensed bosons, the quantum state of these bosons can be effectively represented by a single-particle quantum state. The collective behavior of these interacting bosons, subject to an external potential $V_{ext}$, can be effectively described by the Hamiltonian \cite{Harko}
\begin{multline}\label{31}
\hat{H}=\int d\Vec{r}\hat{\Psi}^{+}(\Vec{r})\left[-\frac{\hbar}{2m}\bigtriangledown^2+V_{rot}\Vec{r}+V_{ext}\Vec{r}\right]\hat{\Psi}(\Vec{r})\\
+\frac{1}{2}\int d\Vec{r} d\Vec{r}^{'}\hat{\Psi}^{+}(\Vec{r})\hat{\Psi}^{+}\Vec{r}^{'}V(\Vec{r}-\Vec{r}^{'})\hat{\Psi}(\Vec{r})\hat{\Psi}(\Vec{r}^{'}),
\end{multline}
where $\hat{\Psi}(\Vec{r})$ and $\hat{\Psi}^{+}(\Vec{r})$ represent the boson field operators responsible for annihilating and creating a particle at the position $\Vec{r}$, respectively. The term $V(\Vec{r}-\Vec{r}^{'})$ denotes the two-body interatomic potential \cite{Dalfovo}. For this paper, we neglect the potential associated with the rotation of the condensate, thus setting $V_{rot}(\Vec{r})=0$. Assuming $V_{ext}(\Vec{r})$ to be the gravitational potential $V$, we satisfy the Poisson equation given by \cite{Harko}
$$\bigtriangledown^2 V=4\pi G \rho_m,$$
Here, $\rho_m=m\rho$ represents the mass density within the BEC. By considering only the first approximation and disregarding the rotation of the BEC, i.e., $V_{rot}=0$, we can calculate the radius R of the BEC as \cite{Harko}
\begin{equation}\label{32}
    R=\pi \sqrt{\frac{\hbar^2\alpha}{Gm^3}},
\end{equation}
Here, $\alpha$, known as the scattering length, is linked to the scattering cross-section of particles within the condensate. Various estimations of the mass and scattering length of condensate dark matter particles exist in the literature. For instance, the scattering length $\alpha$, inferred from astrophysical observations of the Bullet Cluster, falls within the range of $10^{-7}$ $fm$, with the mass of the dark matter particle being approximately $\mu e V$ \cite{Liang}.
The following result is retrieved for the density distribution of the dark matter BEC \cite{Harko}
\begin{equation}\label{33}
    \rho(r)=\rho_s\frac{\sin{kr}}{kr},
\end{equation}
where $\rho_s$ denotes the central density of the condensate and $k=\sqrt{Gm^3/\hbar^2\alpha}$. The mass profile of the dark condensate galactic halo can be read as
\begin{equation}\label{34}
    M(r)=4\pi \int_0^r \rho(r) r^2 dr.
\end{equation}
and its solution is
\begin{equation}
\label{35}
M(r)=\frac{4\pi \rho_s}{k^2}r\left(\frac{\sin{kr}}{kr}-\cos{kr}\right).
\end{equation}
From the above Eq. \eqref{35}, we can obtain the tangential velocity of a test particle moving in the dark halo from the following relation
\begin{equation}\label{ab1}
v_{t}^2(r)=G\,M(r)/r,
\end{equation}
and hence
\begin{equation}\label{36}
v_{t}^2(r)=\frac{4\pi G \rho_s}{k^2}\left(\frac{\sin{kr}}{kr}-\cos{kr}\right),
\end{equation}
where $k=\pi/R$. For $r\rightarrow 0$, we have $v_{t}^2(r)\rightarrow 0$.

\subsubsection{Finding solutions}
\begin{figure*}
\centering 
\epsfig{file=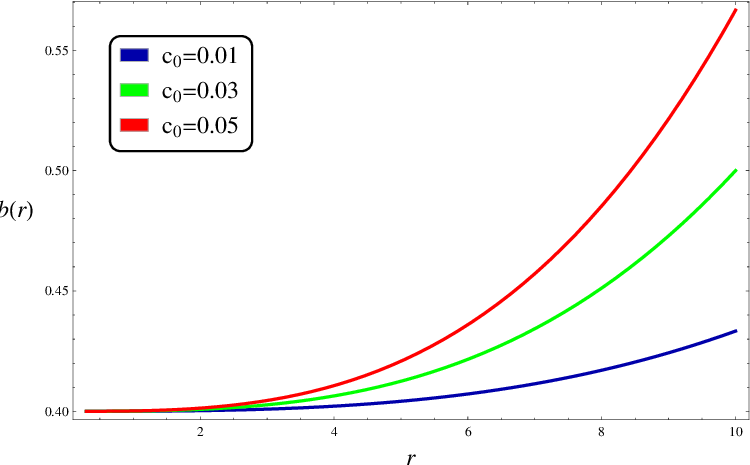, width=.3\linewidth,
height=1.5in} \;\;\;
\epsfig{file=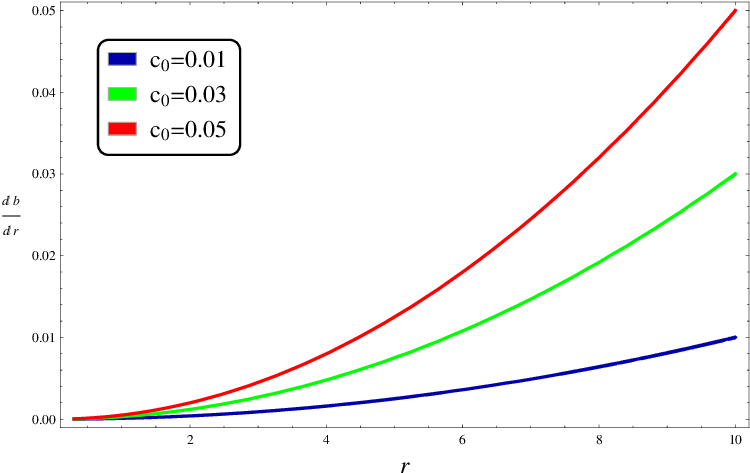, width=.3\linewidth,
height=1.5in} \;\;\;
\epsfig{file=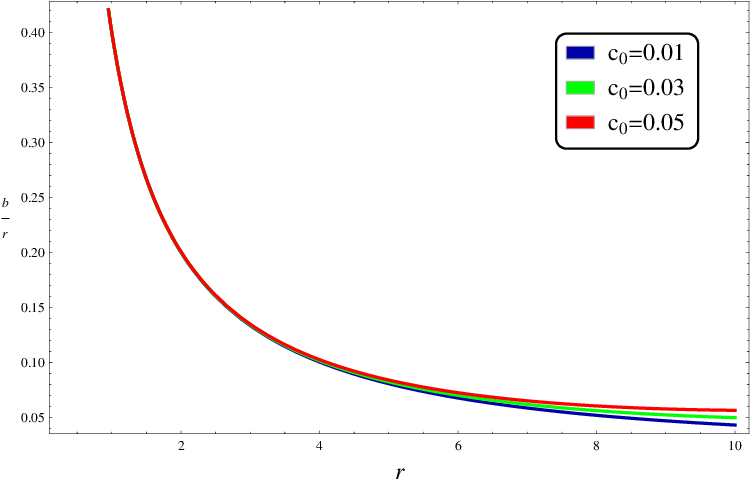, width=.3\linewidth,
height=1.5in}
\caption{\label{F1} Picture represents the behavior of shape functions properties against the radial coordinate $r$ for the BEC model with $c_{1}=0.9$, $r_{0}=0.2$, $\rho_s=0.9$, and $R=1.94$.}
\end{figure*}

Note that within the equatorial plane, the rotational velocity of a test particle in spherically symmetric space-time is defined by \cite{Harko}
\begin{equation}\label{37}
v_{t}^2(r)=r\,\phi^{'}(r).
\end{equation}
Inserting the Eq. \eqref{36} in the above relation \eqref{37}, we have
\begin{equation}\label{38}
\frac{4 R^2 \rho_s}{\pi}\left[\frac{\sin\left({\frac{\pi r}{R}}\right)}{\frac{\pi r}{R}}-\cos\left({\frac{\pi r}{R}}\right)\right]=r\,\phi^{'}(r).
\end{equation}
On solving, we find
\begin{equation}\label{39}
\phi(r)=D_1-\frac{4 \rho_s R^3 \sin \left(\frac{\pi  r}{R}\right)}{\pi ^2 r},
\end{equation}
where $D_1$ is an integrating constant.\\
Now, we equate the density of the wormhole matter with the density of the BEC using Eqs. \eqref{2b9} and \eqref{33}, it follows that


\begin{equation}\label{40}
\frac{c_0 }{2}-\frac{c_1  b'(r)}{r^2}=\frac{\rho_s R}{\pi r}\sin\left({\frac{\pi r}{R}}\right).
\end{equation}
Solving for $b(r)$, we get
\begin{equation}\label{41}
b(r)=\frac{\frac{1}{3} \pi  c_0  r^3-\frac{2 \rho_s R^3 \sin \left(\frac{\pi  r}{R}\right)}{\pi ^2}+\frac{2 \rho_s r R^2 \cos \left(\frac{\pi  r}{R}\right)}{\pi }}{2 \pi  c_1 }+c_3.
\end{equation}
Now to find the integrating constant $c_3$, we impose the throat condition $b(r_0)=r_0$
\begin{multline}\label{42}
c_3=r_0-\frac{1}{2 \pi c_1 } \left[\frac{1}{3} \pi  c_0  r_0^3-\frac{2 \rho_s R^3 \sin \left(\frac{\pi  r_0}{R}\right)}{\pi ^2} \right.\\\left.
+\frac{2 \rho_s R^2 r_0 \cos \left(\frac{\pi  r_0}{R}\right)}{\pi }\right],
\end{multline}
and hence the shape function $b(r)$ becomes
\begin{multline}\label{43}
    b(r)=\frac{1}{6 c_1 }\left[c_0 r^3-c_0  r_0^3+6 c_1  r_0 \right.\\\left.
    +\frac{1}{\pi ^3} \left(6 \rho_s R^2 \left(-R\Lambda_1
    +\pi \Lambda_2\right)\right)\right],
\end{multline}
where
\begin{equation}\label{44}
\Lambda_1=\left(\sin \left(\frac{\pi  r}{R}\right)-\sin \left(\frac{\pi  r_0}{R}\right)\right),
\end{equation}
\begin{equation}\label{45}
\Lambda_2=r \cos \left(\frac{\pi  r}{R}\right)-r_0 \cos \left(\frac{\pi  r_0}{R}\right).
\end{equation}
To sustain the wormhole's structure, the "flaring out" condition must be met to ensure the wormhole mouth is open. The following relation gives this condition at the wormhole throat region
\begin{equation}\label{46}
b^{'}(r_0)=\frac{r_0 \left(\pi  c_0  r_0-2 \rho_s R \sin \left(\frac{\pi  r_0}{R}\right)\right)}{2 \pi  c_1 }<1.
\end{equation}
In Fig. \ref{F1}, we have depicted the flare-out condition for different values of $c_0$, which satisfies the condition around the throat under an asymptotic background. Also, we have checked the behavior of shape function $b(r)$ and noticed that as we increase the value of $c_0$, the shape function increases. Thus, our obtained shape function satisfies all the necessary properties of the shape function. Here, we considered the throat radius $r_0=0.2$.
\subsection{Pseudo isothermal (PI)}
In addition to the BEC-DM model, there is an important class of dark matter models associated with modified gravity, such as Modified Newtonian Dynamics (MOND) \cite{Jusufi3}. In the MOND model, the dark matter density profile is described by the PI profile
\begin{equation}\label{4a1}
\rho_\text{PI}=\frac{\rho_s}{1+\left(\frac{r}{r_s}\right)^2},
\end{equation}
where $\rho_s$ is the central dark matter density and $r_s$ is the scale radius. Unlike the NFW profile, which predicts a sharp density increase (or ``cusp") at the center, the PI profile features a flat central region that better matches the observed rotation curves of low surface brightness and dwarf galaxies. This makes the PI profile especially effective in cases where the NFW model falls short of accurately describing galaxy dynamics \cite{Jusufi3}. Many studies have highlighted the success of the PI profile in matching observed galactic data. For example, de Blok et al. \cite{Frenk4} compared the PI and NFW profiles and found that the PI model's flat core is a better fit for the observed density profiles of low surface brightness galaxies, addressing the "core-cusp problem." Likewise, Gentile et al. \cite{pi1} analyzed spiral galaxy rotation curves and found that the PI profile consistently fits observed data across various radii, further confirming its applicability. The PI profile has also been effective in modeling dwarf galaxies, as demonstrated by Oh et al. \cite{pi2}, who used it to accurately represent the dark matter distribution in galaxies from the THINGS survey without requiring additional adjustments. In the context of wormhole geometry, the PI profile has been examined in \cite{pi3}, and it was found that the dark matter density for an axisymmetric traversable wormhole is similar to that of a black hole spike. However, the dark matter density varies with the wormhole's spin in the opposite direction. This versatility emphasizes the potential of the PI profile in both observational and theoretical physics.

\subsubsection{Finding solution}
\begin{figure*}
\centering 
\epsfig{file=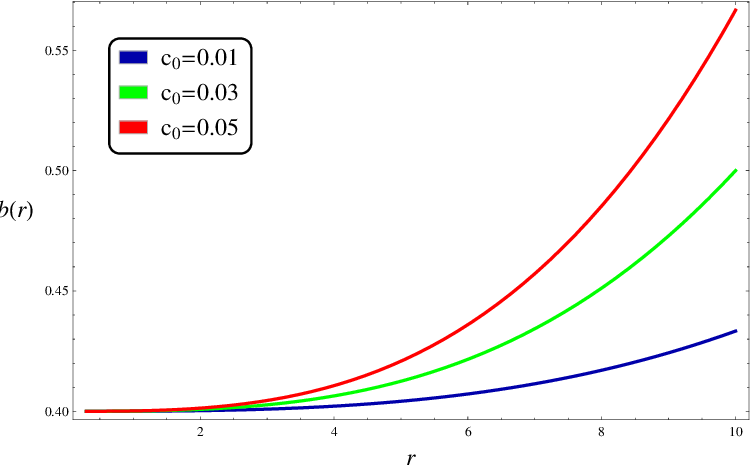, width=.3\linewidth,
height=1.5in}\;\;\; 
\epsfig{file=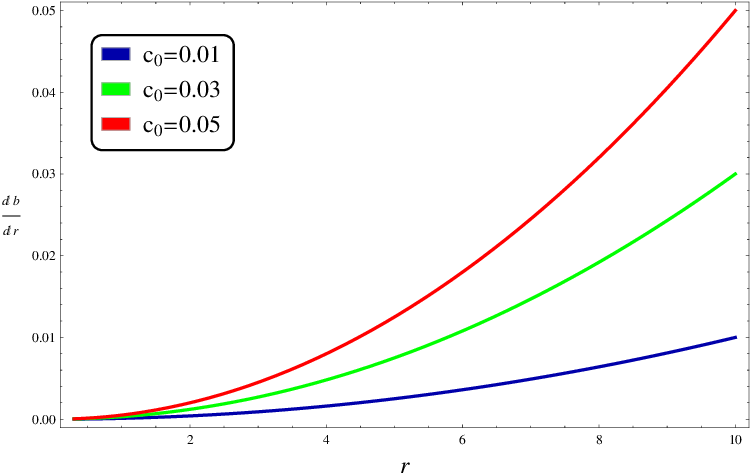, width=.3\linewidth,
height=1.5in}
\;\;\; \epsfig{file=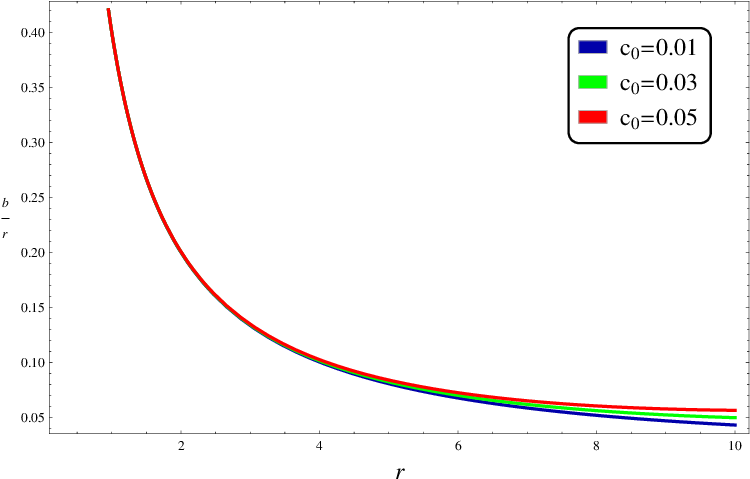,width=.3\linewidth,height=1.5in}
\caption{\label{F11} Picture represents the behavior of shape functions properties against the radial coordinate $r$ for the PI model with $c_{1}=0.9$, $r_{0}=0.2$, $\rho_s=0.9$, and $r_{s}=1.94$.}
\end{figure*}

The solution of mass profile for the PI galactic halo can be obtained using Eq. \eqref{34}
\begin{equation}\label{4a2}
M(r)=4 \pi \rho_s r_s^2 \left(r-r_s \tan ^{-1}\left(\frac{r}{r_s}\right)\right).
\end{equation}
The tangential velocity $v_t^2(r)$ of a test particle moving in the dark halo is given by from Eq. \eqref{ab1}
\begin{equation}\label{4a3}
v_t^2(r)=\frac{4 \pi \rho_s r_s^2}{r} \left(r-r_s \tan ^{-1}\left(\frac{r}{r_s}\right)\right).
\end{equation}
Now, with the relation given in Eq. \eqref{37}, we can find the redshift function
\begin{equation}\label{4a4}
\phi(r)=2 \pi  \rho_s r_s^2 \log \left(r^2+r_s^2\right)+\frac{4 \pi  \rho_s r_s^3 \tan ^{-1}\left(\frac{r}{r_s}\right)}{r}+D_2,
\end{equation}
where $D_2$ is the integrating constant. Now, we shall try to calculate the shape function of the wormhole under the PI dark matter model by comparing the density of the wormhole with the density of PI dark matter. From Eqs. \eqref{2b9} and \eqref{4a1}, one can get the relation
\begin{equation}\label{47}
\frac{c_0 }{2}-\frac{c_1  b'(r)}{r^2}=\frac{\rho_s}{1+\left(\frac{r}{r_s}\right)^2}.
\end{equation}
On solving
\begin{equation}\label{48}
b(r)=\frac{\frac{c_0  r^3}{3}+2 \rho_s r_s^3 \tan ^{-1}\left(\frac{r}{r_s}\right)-2 \rho_s r r_s^2}{2 c_1 }+c_5,
\end{equation}
where $c_5$ is the integrating constant. By imposing the throat condition on the above equation, we can obtain the final shape function
\begin{multline}\label{49}
b(r)=\frac{1}{6 c_1 }\left[c_0  r^3+6 \rho_s r_s^3 \Lambda_4
+6 \rho_s r_s^2 (r_0-r) \right.\\\left.
-c_0 r_0^3+6 c_1  r_0\right].
\end{multline}
where
\begin{equation}\label{50}
\Lambda_4=\left(\tan ^{-1}\left(\frac{r}{r_s}\right)-\tan ^{-1}\left(\frac{r_0}{r_s}\right)\right).
\end{equation}
Now, we check the flare-out condition at the throat, i.e., 
\begin{equation}
b'(r_0)= \frac{\text{r1}^2 r_0^2 (c_0 -2 \rho_s)+c_0  r_0^4}{2 c_1  \left(r_s^2+r_0^2\right)},
\end{equation}
which is obviously satisfied. For more information, one can check Fig \ref{F11}.
\subsection{Navarro-Frenk-White (NFW)}
An approximate analytical formulation of the NFW density profile is established by drawing upon the Cold Dark Matter ($\Lambda$CDM) theory and numerical simulations \cite{Rahaman3,Frenk3}. In the context of galaxies and clusters, the dark matter halo can be characterized through the NFW density profile, expressed as
\begin{equation}\label{51}
\rho(r)=\frac{\rho_s}{(r/r_s)(1+r/r_s)^2},
\end{equation}
where $\rho_s$ represents the characteristic density and $r_s$ represents the scale radius that distinguishes the transition between the inner and outer regions of the halo. Near the center, the density scales as $\rho(r) \propto r^{-1}$, forming a central ``cusp", while further out, it decreases more sharply as $\rho(r) \propto r^{-3}$ \cite{Rahaman3}. This profile has proven effective across a broad range of halo sizes, from small dwarf galaxies to large galaxy clusters, demonstrating its wide applicability. However, a key limitation, known as the ``core-cusp problem", arises because the NFW model predicts a steep central density that does not align with the flatter cores observed in some galaxies, particularly those with lower mass \cite{Frenk4}.
Further investigations have continued to refine and challenge the NFW profile. For example,  Navarro et al. \cite{Frenk3} improved the model through more detailed simulations, affirming its validity but also highlighting some inconsistencies in the inner regions of halos. These discrepancies have prompted the development of alternative profiles, such as the Einasto profile, which introduces an extra parameter that allows more flexibility in the shape of the inner halo \cite{Frenk5}. Additionally, research has shown that baryonic processes, including supernova feedback and the influence of active galactic nuclei, can alter the density profile, smoothing the central region and bringing model predictions closer to observed data \cite{Frenk6}. Despite these issues, the NFW profile remains a useful and commonly applied model, particularly for smaller halos like dwarf galaxies, due to its relative simplicity \cite{Frenk8}.
\subsubsection{Finding solution}
\begin{figure*}
\centering 
\epsfig{file=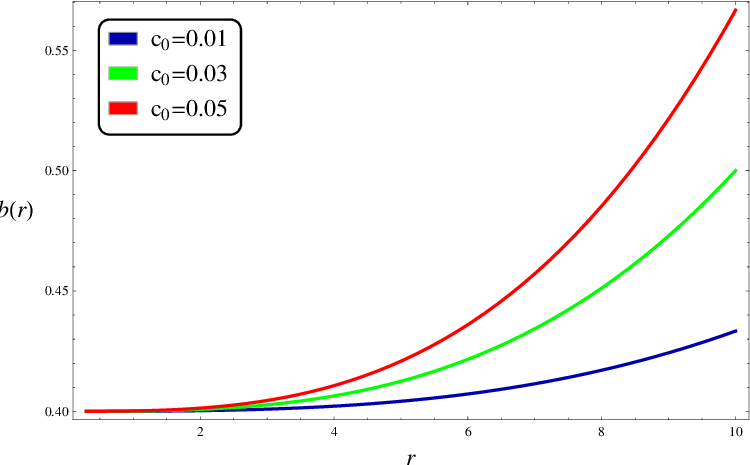, width=.3\linewidth,
height=1.5in} \;\;\;
\epsfig{file=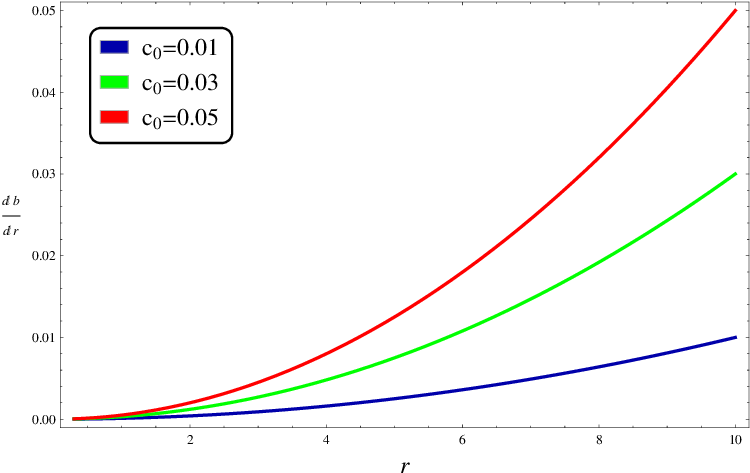, width=.3\linewidth,
height=1.5in}
\;\;\; \epsfig{file=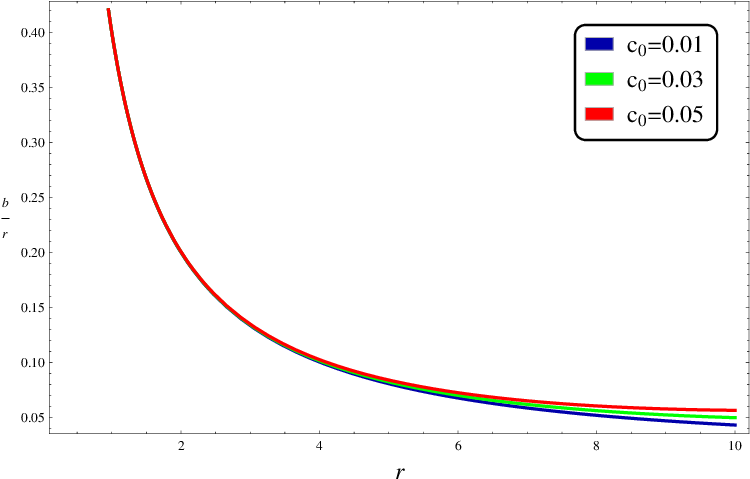,width=.3\linewidth, height=1.5in}
\caption{\label{F111} Picture represents the behavior of shape functions properties against the radial coordinate $r$ for the NFW model with $c_{1}=0.9$, $r_{0}=0.2$, $\rho_s=0.9$, and $r_{s}=1.94$.}
\end{figure*}

For this case, the mass function can be read as
\begin{equation}\label{4a2}
M(r)=4 \pi \rho_s r_s^3 \left(\frac{r_s}{r+r_s}+\log (r+r_s)-\log (r_s)-1\right),
\end{equation}
and consequently, the tangential velocity $v_t^2(r)$ is
\begin{equation}\label{52}
v_t^2(r)=\frac{4 \pi \rho_s r_s^3}{r} \left(\frac{r_s}{r+r_s}+\log (r+r_s)-\log (r_s)-1\right).
\end{equation}
Now from Eq. \eqref{37}, we can find the redshift function
\begin{equation}\label{53}
\phi(r)=\frac{4 \pi \rho_s r_s^3}{r}\left(\log (r_s)-\log (r+r_s)\right)+D_3,
\end{equation}
where $D_3$ is the integrating constant.
Again, from Eqs. \eqref{2b9} and \eqref{51}, it follows that
\begin{equation}\label{54}
\frac{c_0 }{2}-\frac{c_1  b'(r)}{r^2}=\frac{\rho_s}{(r/r_s)(1+r/r_s)^2}
\end{equation}
On solving the above differential equation, we obtain
\begin{multline}\label{55}
b(r)=\frac{1}{2 c_1 }\left[-\frac{2 \rho_s r_s^4}{r+r_s}-2 \rho_s r_s^3 \log (r+r_s)+c_0  r_s^2 (r+r_s)\right.\\\left.
-c_0  r_s (r+r_s)^2+\frac{1}{3} c_0  (r+r_s)^3\right]+c_7,
\end{multline}
where $c_7$ is the integrating constant. Now, we impose the condition $b(r_0)=r_0$ on the above equation and obtain the shape function
\begin{multline}\label{56}
b(r)=\frac{1}{6 c_1 }\left[c_0  r^3+\Lambda_6+6 \rho_s r_s^3 \Lambda_7-c_0  r_0^3+6 c_1  r_0\right],
\end{multline}
where,
\begin{equation}\label{57}
\Lambda_6=\frac{6 \rho_s r_s^4 (r-r_0)}{(r+r_s) (r_s+r_0)},   
\end{equation}
\begin{equation}\label{58}
\Lambda_7=(\log (r_s+r_0)-\log (r+r_s)).
\end{equation}
Now, we study the important condition, i.e., flare-out condition, with the obtained shape function \eqref{56} graphically. We noticed that with the appropriate choices of parameters, the flare-out condition is satisfied, i.e., $b(r)<1$ at $r=r_0$ (see Fig. \eqref{F111}).

\begin{figure*}
\centering \epsfig{file=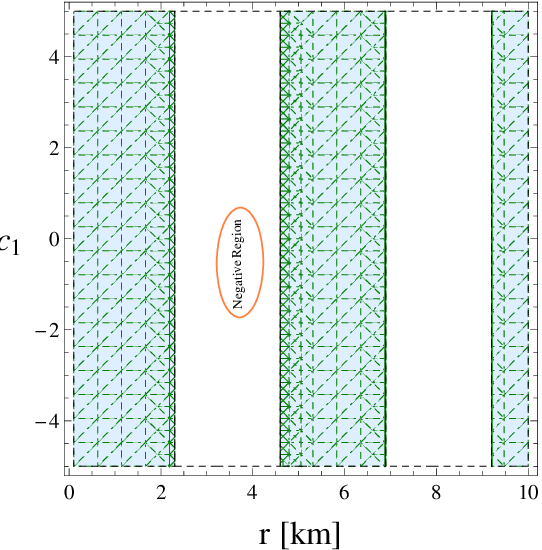, width=.32\linewidth,
height=1.5in}\;\;\; \epsfig{file=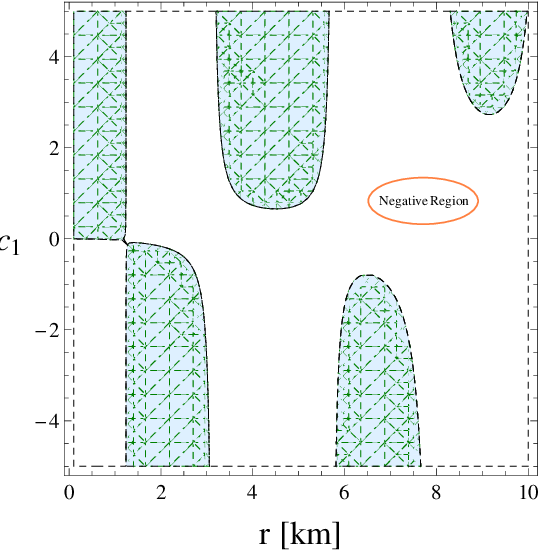, width=.32\linewidth,
height=1.5in}\;\;\;\epsfig{file=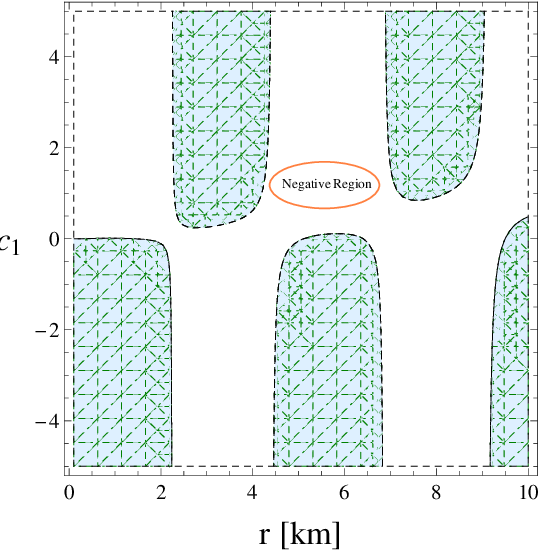, width=.32\linewidth,
height=1.5in}
\centering \epsfig{file=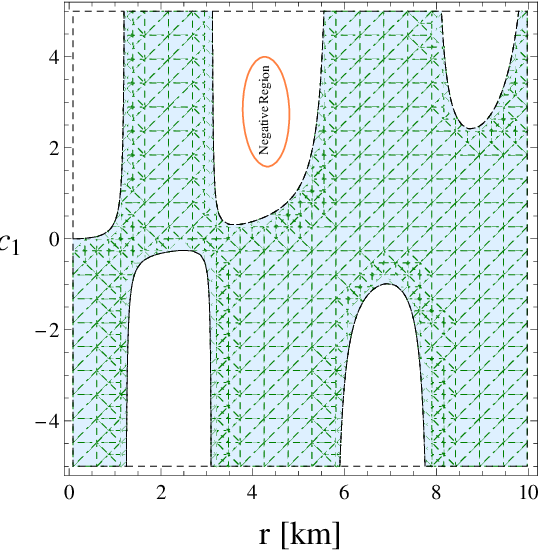, width=.32\linewidth,
height=1.5in}\;\;\; \epsfig{file=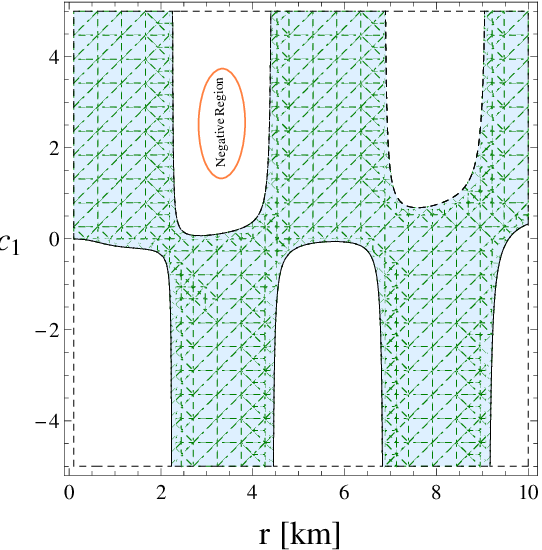, width=.32\linewidth,
height=1.5in}\;\;\;\epsfig{file=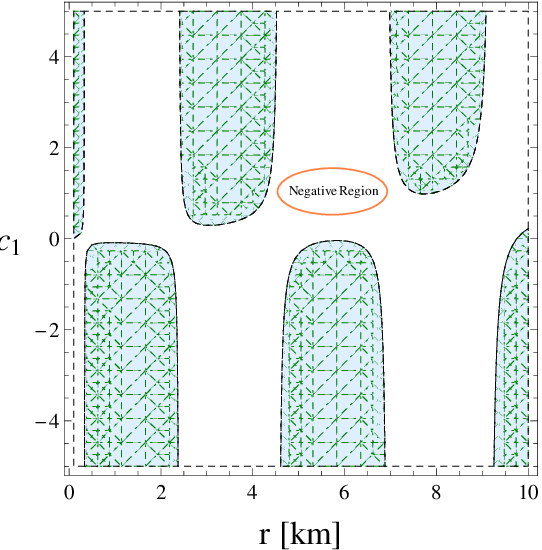, width=.32\linewidth,
height=1.5in}
\caption{\label{F5} Picture represents the valid and negative regions of all the energy conditions for the BEC profile. In the first row $\rho$ (left), $\rho+P_r$ (middle), and $\rho+P_t$ (right) are presented. In second row, $\rho-\abs{P_r}$ (left), $\rho-\abs{P_t}$ (middle), and $\rho+P_r +2P_t$ (right) are shown with $c_{0}=0.05$, $r_{0}=0.2$, $\rho_{c}=0.9$, and $R=1.94$.}
\end{figure*}

\begin{figure*}
\centering \epsfig{file=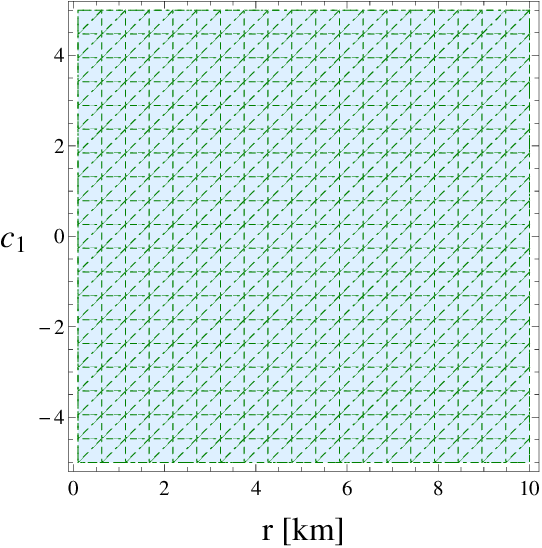, width=.32\linewidth,
height=1.5in}\;\;\; \epsfig{file=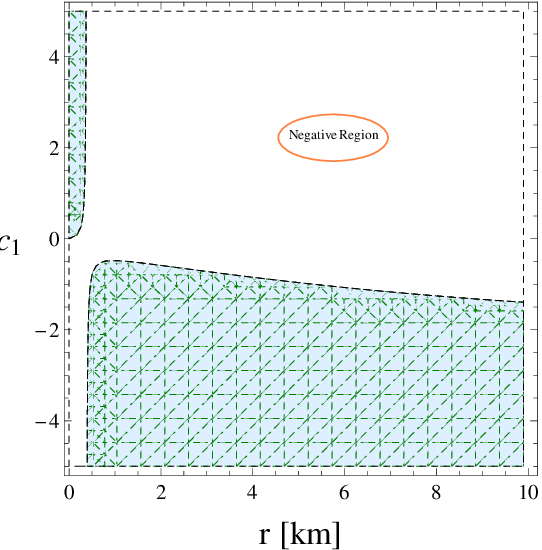, width=.32\linewidth,
height=1.5in}\;\;\;\epsfig{file=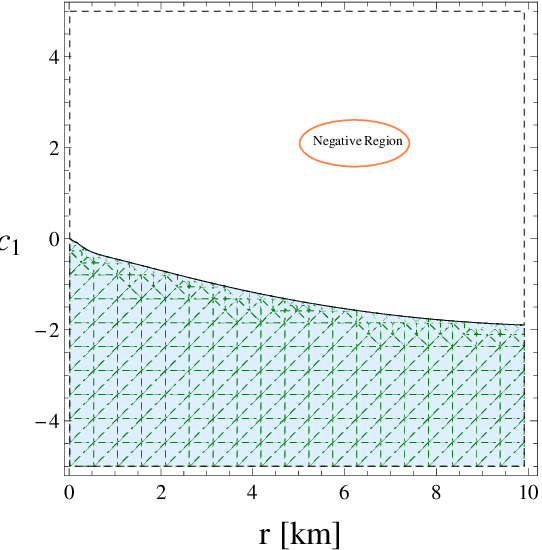, width=.32\linewidth,
height=1.5in}
\centering \epsfig{file=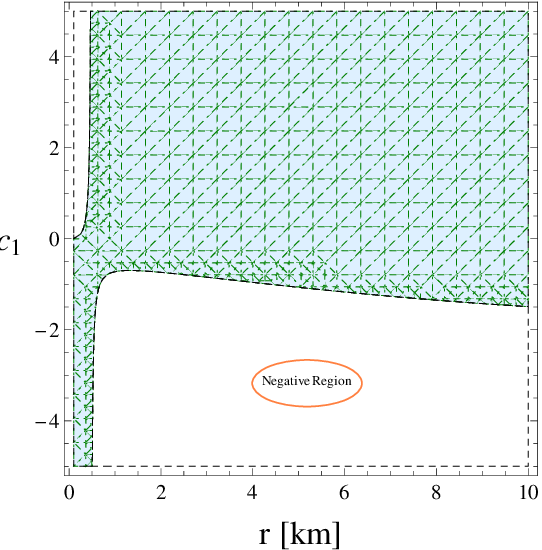, width=.32\linewidth,
height=1.5in}\;\;\; \epsfig{file=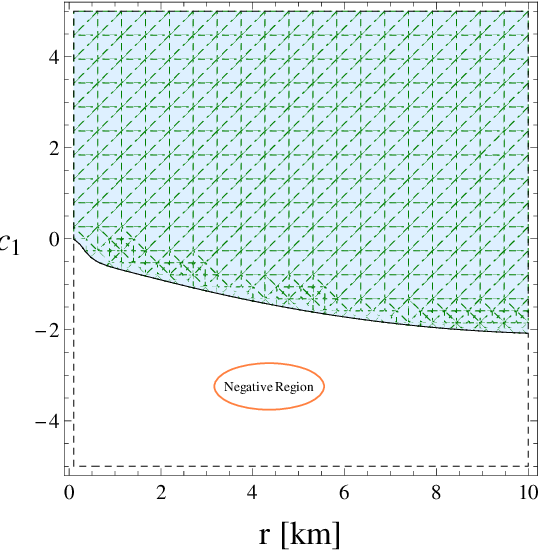, width=.32\linewidth,
height=1.5in}\;\;\;\epsfig{file=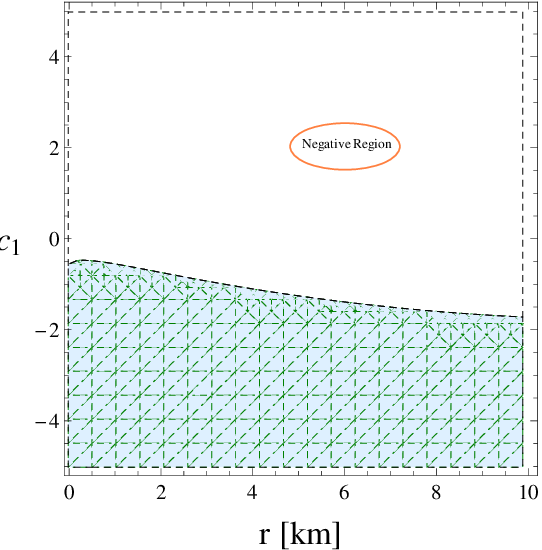, width=.32\linewidth,
height=1.5in}
\caption{\label{F6} Picture represents the valid and negative regions of all the energy conditions for the PI profile. In the first row $\rho$ (left), $\rho+P_r$ (middle), and $\rho+P_t$ (right) are presented. In second row, $\rho-\abs{P_r}$ (left), $\rho-\abs{P_t}$ (middle), and $\rho+P_r +2P_t$ (right) are shown with $c_{0}=0.05$, $r_{0}=0.2$, $\rho_s=0.9$, and $r_{s}=1.94$.}
\end{figure*}

\begin{figure*}
\centering \epsfig{file=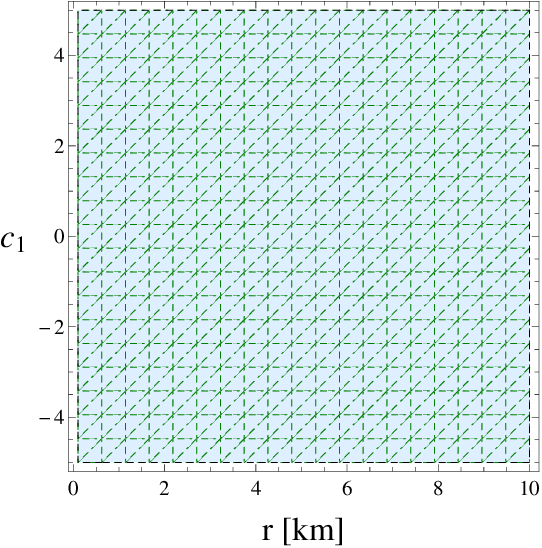, width=.32\linewidth,
height=1.5in}\;\;\; \epsfig{file=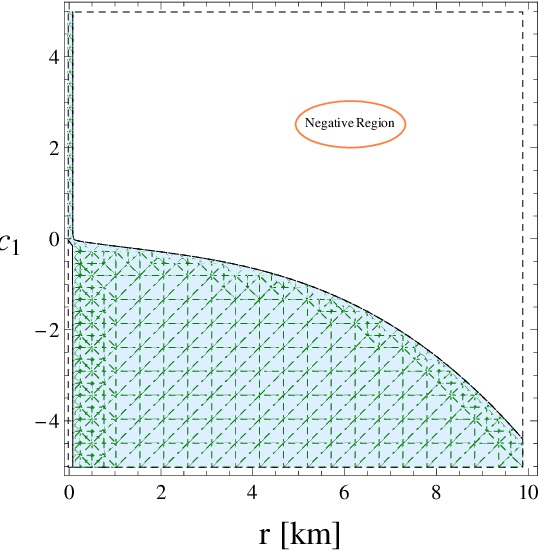, width=.32\linewidth,
height=1.5in}\;\;\;\epsfig{file=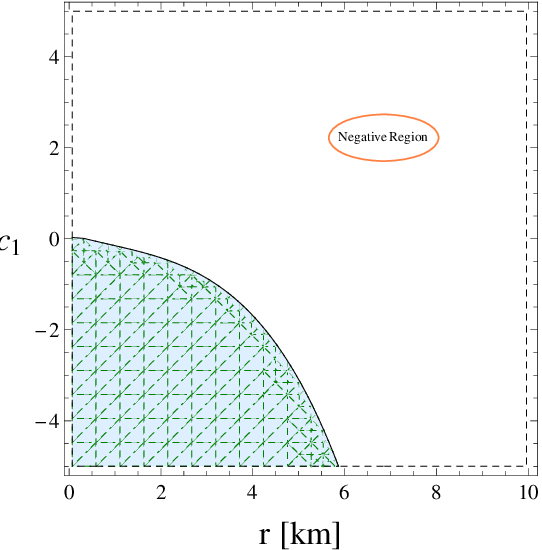, width=.32\linewidth,
height=1.5in}
\centering \epsfig{file=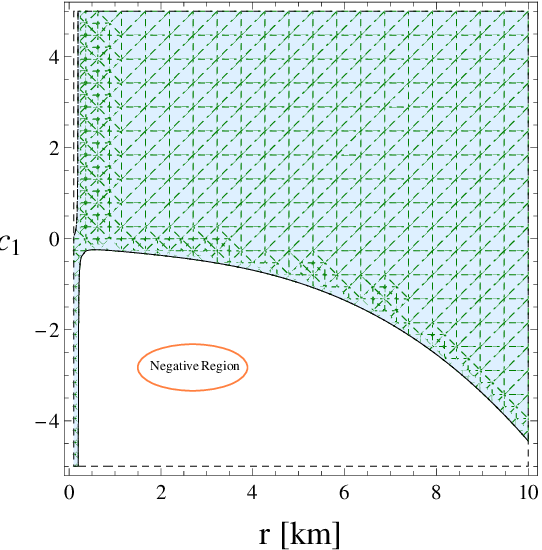, width=.32\linewidth,
height=1.5in}\;\;\; \epsfig{file=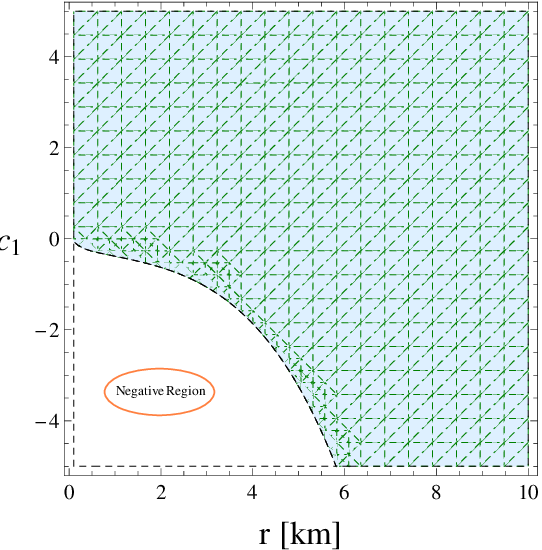, width=.32\linewidth,
height=1.5in}\;\;\;\epsfig{file=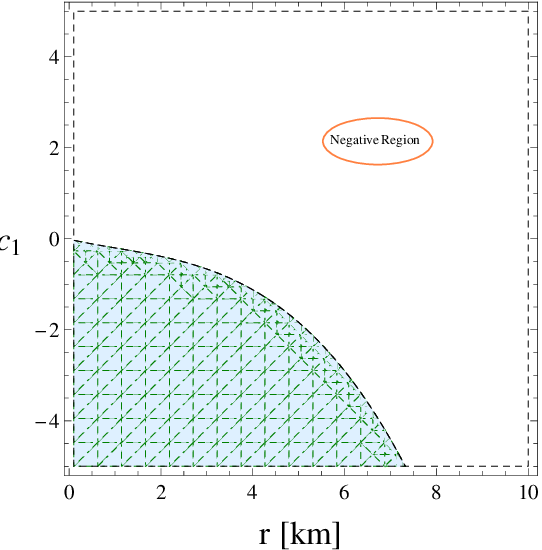, width=.32\linewidth,
height=1.5in}
\caption{\label{F7} Picture represents the valid and negative regions of all the energy conditions for the NFW profile. In the first row $\rho$ (left), $\rho+P_r$ (middle), and $\rho+P_t$ (right) are presented. In second row, $\rho-\abs{P_r}$ (left), $\rho-\abs{P_t}$ (middle), and $\rho+P_r +2P_t$ (right) are shown with $c_{0}=0.05$, $r_{0}=0.2$, $\rho_s=0.9$, and $r_{s}=1.94$.}
\end{figure*}
\section{Energy conditions}\label{sec4}
In the study of modified gravitational theories, the validity of energy conditions of matter is often the key issue, and the ECs are the necessary conditions to explain the singularity theorem \cite{Hawking}. Moreover, the ECs aid in analyzing the entire space-time structure without precise solutions to Einstein's equations and play a crucial role in investigating wormhole solutions in the context of modified gravity theories. The common popular basic energy conditions (the null, weak, dominant, and strong energy conditions) originate from the Raychaudhuri equation \cite{Raychaudhuri}, which plays a crucial role in describing the attractive properties of gravity and positive energy density. The WEC \cite{Visser} is defined by $T_{\mu\nu}U^{\mu}U^{\nu}\geq 0$ i.e.,
\begin{equation}
    \rho \geq 0, \quad \rho+P_r\geq 0, \quad \rho+P_t\geq 0,
\end{equation}
where $U^{\mu}$ denotes the time-like vector. This means that local energy density is positive, and it gives rise to the continuity of NEC, which is defined by $T_{\mu\nu}k^{\mu}k^{\nu}\geq 0$, i.e., 
\begin{equation}
    \rho+P_r\geq 0, \quad \rho+P_t\geq 0,
\end{equation}
where $k^{\mu}$ represents a null vector. On the other hand, strong energy condition (SEC) stipulates that
\begin{equation}
\rho+P_r\geq 0, \quad \rho+P_t\geq 0,\quad \text{and} \quad \rho+P_r+2P_t\geq 0
\end{equation}
and the dominant energy conditions (DEC) are defined by
\begin{equation}
\rho \geq 0, \quad \rho-\abs{P_r}\geq 0, \quad \rho-\abs{P_t}\geq 0.
\end{equation}
Now, with the above expressions, we check the behaviors of ECs.\\
The expression for radial NEC for the BEC DM profile can be read as
\begin{multline}\label{59}
\rho+P_r=\frac{1}{6} \left[-3 c_0 +\frac{1}{r^3}\left(c_0  r^3+\frac{1}{\pi ^3}\left(6 \rho_s R^2 \right.\right.\right.\\\left.\left.\left.
\times \left(\pi \Lambda_2-R \Lambda_1\right)\right)-c_0  r_0^3+6 c_1  r_0\right)-\frac{1}{\pi ^5 r^4}\left(8 \rho_s R^2 \right.\right.\\\left.\left.
\times \Lambda_3 \left(\pi ^3 \left(-c_0  r^3+6 c_1  r +c_0  r_0^3-6 c_1  r_0\right)+6 \rho_s R^2 \right.\right.\right.\\\left.\left.\left.
\times \left(R \Lambda_1 -\pi \Lambda_2\right)\right)\right)\right]+\frac{\rho_s R \sin \left(\frac{\pi  r}{R}\right)}{\pi  r},
\end{multline}
where $\Lambda_3=\left(R \sin \left(\frac{\pi  r}{R}\right)-\pi  r \cos \left(\frac{\pi  r}{R}\right)\right)$. $\Lambda_1$ and $\Lambda_2$ are defined in \eqref{44} and \eqref{45}.\\
Similarly, for the PI case, NEC can be obtained as follows:
\begin{multline}\label{60}
\rho+P_r=\frac{\rho_s r_s^2}{r^2+r_s^2}+\frac{1}{6 r^4}\left[-3 c_0  r^4+r \left(c_0  r^3+6 \rho_s r_s^3 \Lambda_4 \right.\right.\\\left.\left.
+6 \rho_s r_s^2 (r_0-r)-c_0  r_0^3+6 c_1  r_0\right)-8 \pi \rho_s r_s^2 \Lambda_5 \left(-c_0  r^3 +6 c_1  r \right.\right.\\\left.\left.
-6 \rho_s r_s^3 \Lambda_4+6 \rho_s r_s^2 (r-r_0)+c_0  r_0^3-6 c_1  r_0\right)\right],
\end{multline}
where, $\Lambda_4$ is already defined in \eqref{50} and $\Lambda_5=\left(r-r_s \tan ^{-1}\left(\frac{r}{r_s}\right)\right)$.\\
At last, the expression for NEC for the NFW profile
\begin{multline}\label{61}
\rho+P_r=-\frac{c_0 }{2}+\frac{\Lambda_6+c_0  r^3+6 \Lambda_7 \rho_s r_s^3-c_0  r_0^3+6 c_1  r_0}{6 r^3}\\
-\frac{1}{r^4 (r+r_s)}\left[8 \pi  c_1  \rho_s r_s^3 ((r+r_s) \Lambda_8+r) \left(\frac{1}{6 c_1 }\left(\Lambda_6+c_0  r^3 \right.\right.\right.\\\left.\left.\left.
+6 \Lambda_7 \rho_s r_s^3-c_0  r_0^3+6 c_1 r_0\right)-r\right)\right]+\frac{\rho_s r_s^3}{r (r+r_s)^2},
\end{multline}
where $\Lambda_8=(\log (r_s)-\log (r+r_s))$. The expression of $\Lambda_6$ and $\Lambda_7$ are presented in \eqref{57} and \eqref{58}, respectively.\\
Now, at the throat ($r=r_0$), the NEC for each DM halo profile has been obtained and shown in Eq. \eqref{62}.
\begin{equation}\label{62}
\rho+P_r\bigg\vert_{r=r_0}=
     \begin{cases}
      -\frac{c_0 }{2}+\frac{\rho_s R \sin \left(\frac{\pi  r_0}{R}\right)}{\pi  r_0}+\frac{c_1 }{r_0^2},  & \text{(BEC)}\\
      \\
      -\frac{c_0 }{2}+\frac{\rho_s r_s^2}{r_s^2+r_0^2}+\frac{c_1 }{r_0^2}, &  \text{(PI)}\\
      \\
       \frac{1}{r_0^2}\left[c_1 +\frac{\rho_s r_s^3 r_0}{(r_s+r_0)^2}\right]-\frac{c_0 }{2},   &  \text{(NFW)}
     \end{cases}
\end{equation}
\subsection{Detailed analysis of these DM models with graphical descriptions}
The primary focus of research has been on a method where the energy conditions are not violated by the actual matter itself but rather by an effective energy-momentum tensor that emerges within a modified gravitational theory framework. In this section, we will examine the energy conditions for the solutions we have explored. A comprehensive graphical analysis, including a regional investigation of the energy conditions in these dark matter models, is presented in Figs. (\ref{F5}-\ref{F7}). The energy density $\rho$ for the BEC model shows positive behavior in the vicinity of the throat within $-5 \leq c_1 \leq 5$ and $0 \leq r \leq 2$. Interestingly, for the PI and NFW cases, energy density shows positive behavior in the entire space-time. Next, we plotted the radial NEC $\rho+P_r$ graphs for each case. It was noticed that for the BEC case, $\rho+P_r$ is violated near the throat for $-5\leq c_1 \leq 0$ and $0\leq r \leq 1.7$, and satisfied for $0\leq c_1 \leq 5$. For the PI case, the valid region of $\rho+P_r$ is $-5\leq c_1 < -0.5$ and negative region is $-0.5\leq c_1 < 5$ against the radial co-ordinate $0.4 \leq r \leq10$. The energy condition $\rho+P_r$ for the NFW profile depicted violated for $0\leq c_1 \leq 5$ and obeyed for $-5\leq c_1 \leq 0$ within $0\leq r \leq 10$. Further, $\rho+P_t$ is depicted for each case and observed that it is disrespected for the BEC case within $0\leq c_1 \leq 5$ and satisfied within $-5 \leq c_1 \leq 0$ against $0\leq r \leq 2.2$. Also, for the PI case, $\rho+P_t$ shows negative behavior for $c_1\geq 0$ whereas satisfied within $c_1<0$. In addition, $\rho+P_t$ for the NFW case portrays a valid region for $-5\leq c_1 \leq 0$ and $0\leq r \leq 6$, and the remaining region shows the violation of $\rho+P_t$. Furthermore, we have thoroughly investigated the behavior of SEC, which can be found in the lower left plot of figures (\ref{F5}-\ref{F7}). We observed that the negative region of $\rho+P_r+2P_t$ for the BEC case is $0\leq c_1 \leq 5$, PI case is $c_1>-0.5$, and for the NFW case is $c_1\geq 0$. Finally, we checked the behavior of DEC for each profile, and interestingly, we noticed that DEC's behavior was completely opposite to NEC's behavior. One can check figs. (\ref{F5}-\ref{F7}) for a complete overview. It is important to note that the model parameter, $c_1$, significantly influences all the energy conditions in the current analysis. In fact, within the range of parameters involved, all the energy conditions are violated in maximum regions, confirming the presence of exotic matter, which is necessary for creating the obtained wormhole solutions due to the violation of energy conditions, specifically the violation of NEC. This exotic matter is believed to contribute to the stability and traversability of wormholes by counteracting the gravitational collapse caused by ordinary matter. The study of energy violation in the context of wormholes sheds light on these structures. The energy conditions for each case (as shown in Figs. (\ref{F5}-\ref{F7})) support the existence of these wormhole solutions in the background of symmetric teleparallel gravity.\\
All the results mentioned above regarding the energy conditions for each DM model are also summarized in Table-\ref{Table1}

\begin{table*}[t]
    \centering
\begin{tabular}{ p{3cm}| p{12cm}}
 \hline
  \multicolumn{2}{|c|}{The behavior of the energy conditions around the throat} \\
 \hline
Physical expressions & \quad \quad \quad \quad \quad \quad \quad \quad \quad \quad \quad BEC profile\\
 \hline
$\rho$ & $\rho>0$ for $-5\leq c_1 \leq 5$\\
\hline
$\rho+P_r$ & $\rho+P_r<0$ for $-5\leq c_1 \leq 0$ and $\rho+P_r>0$ for $0\leq c_1 \leq 5$\\
\hline
$\rho+P_t$ & $\rho+P_t<0$ for $0\leq c_1 \leq 5$ and $\rho+P_t>0$ for $-5\leq c_1 \leq 0$\\
\hline
$\rho+P_r+2P_t$ & $\rho+P_r+2P_t<0$ for $0\leq c_1 \leq 5$ and $\rho+P_r+2P_t>0$ for $-5\leq c_1 \leq 0$\\
\hline
$\rho-\abs{P_r}$ & $\rho-\abs{P_r}<0$ for $0\leq c_1 \leq 5$ and $\rho-\abs{P_r}>0$ for $-5\leq c_1 \leq 0$\\
\hline
$\rho-\abs{P_t}$ & $\rho-\abs{P_t}<0$ for $-5\leq c_1 \leq 0$ and $\rho-\abs{P_t}>0$ for $0\leq c_1 \leq 5$\\
\hline
\multicolumn{2}{|c|} {PI profile}\\
\hline
$\rho$ & $\rho>0$ for $-5\leq c_1 \leq 5$\\
\hline
$\rho+P_r$ & $\rho+P_r<0$ for $-0.5\leq c_1 < 5$ and $\rho+P_r>0$ for $-5\leq c_1<-0.5$\\
\hline
$\rho+P_t$ & $\rho+P_t<0$ for $c_1 \geq 0$ and $\rho+P_t>0$ for $c_1<0$\\
\hline
$\rho+P_r+2P_t$ & $\rho+P_r+2P_t<0$ for $c_1 >-0.5$ and $\rho+P_r+2P_t>0$ for $c_1 \leq -0.5$\\
\hline
$\rho-\abs{P_r}$ & $\rho-\abs{P_r}<0$ for $-5\leq c_1<-0.5$ and $\rho-\abs{P_r}>0$ for $-0.5\leq c_1 < 5$\\
\hline
$\rho-\abs{P_t}$ & $\rho-\abs{P_t}<0$ for $c_1<0$ and $\rho-\abs{P_t}>0$ for $c_1 \geq 0$\\
\hline
\multicolumn{2}{|c|} {NFW profile}\\
\hline
$\rho$ & $\rho>0$ for $-5\leq c_1 \leq 5$\\
\hline
$\rho+P_r$ & $\rho+P_r<0$ for $0\leq c_1 < 5$ and $\rho+P_r>0$ for $-5\leq c_1<0$\\
\hline
$\rho+P_t$ & $\rho+P_t<0$ for $c_1 \geq 0$ and $\rho+P_t>0$ for $-5\leq c_1<0$\\
\hline
$\rho+P_r+2P_t$ & $\rho+P_r+2P_t<0$ for $c_1 \geq 0$ and $\rho+P_r+2P_t>0$ for $0< c_1 \leq -5$\\
\hline
$\rho-\abs{P_r}$ & $\rho-\abs{P_r}<0$ for $-5\leq c_1<0$ and $\rho-\abs{P_r}>0$ for $0\leq c_1 \geq 5$\\
\hline
$\rho-\abs{P_t}$ & $\rho-\abs{P_t}<0$ for $-5\leq c_1<0$ and $\rho-\abs{P_t}>0$ for $c_1 \geq 0$\\
\hline
\end{tabular}
\caption{Outlook of the energy conditions}
\label{Table1}
\end{table*}
\section{Shadows of Wormholes}\label{sec5} 
In this section, we shall discuss wormhole shadows under the effect of three different kinds of dark matter halos, including BEC. 
To study the impact of wormholes on light deflection, we need to calculate the movement of light rays. We consider the motion of a light beam using the null geodesic equation, which allows us to predict its trajectory. The equation can be found by applying the Euler-Lagrange equation:
$\mathcal{L}=-\frac{1}{2}g_{\mu \nu}\dot{x}^{\mu}\dot{x}^{\nu}$.
Without loss of generality, we can consider the equatorial plane, i.e., $\theta=\frac{\pi}{2}$. The Lagrangian equation describing the motion of light rays around the wormhole geometry is given as
\begin{equation}\label{rrr2}
\begin{split}
 & \mathcal{L}=-\frac{1}{2}g_{\mu \nu}\dot{x}^{\mu}\dot{x}^{\nu}\\
& =- e^{2\phi(r)} \dot{t}^2+\frac{1}{1-\frac{b(r)}{r}} \dot{r}^2 +r^2 (\dot{\theta}^2 + \sin^2\theta \dot{\Phi}^2),\\
\end{split}
\end{equation}
where $\dot{x}^{\mu}$ denotes the four-velocity of the photon, and the dot represents the differentiation with respect to the affine parameter $\tau$. Now, by applying the Lagrangian equation of motion within the scope of wormhole geometry in Eq. (\ref{rrr2}), one can get the following relations
\begin{equation}\label{rrr3}
\dot{t}=\frac{dt}{d\tau}=\frac{E}{e^{2\phi(r)}},
\end{equation}
\begin{equation}\label{rrr4}
\dot{\Phi}= \frac{d\Phi}{d\tau}=\frac{L}{r^2 \sin^2(\theta)}.
\end{equation}
In the above relations, $E$ and $L$ represent the energy and angular momentum of the particle around the wormhole throat. To simplify the geodesics, we propose two dimensionless impact parameters: $\xi=\frac{L}{E}$ and $\eta=\frac{\kappa}{E^2}$, where $\kappa$ denotes the Carter constant. By considering the null geodesic ($\mathcal{L}=0$), which can be expressed in terms of kinetic and potential energy. Scale an affine parameter $\tau \rightarrow \frac{\tau}{L}$. The orbit equation of motion can be revised as:
\begin{equation}\label{rrr6}
   K_{E}+V_{eff}=\frac{1}{\Tilde{b}^2},
\end{equation}
where  $\Tilde{b}=\frac{L}{E}$ is the  kinetic energy function $K_{E}$ and potential function $V_{eff}$ are described as
\begin{equation}\label{rrr7}
 K_{E}=  \frac{e^{2\phi(r)}}{1-\frac{b(r)}{r}}\dot{r}^2,
\end{equation}
and
\begin{equation}\label{rrr7}
 V_{eff}=   \frac{e^{2\phi(r)}}{r^2}.
\end{equation}
\par
\begin{figure*}
\centering \epsfig{file=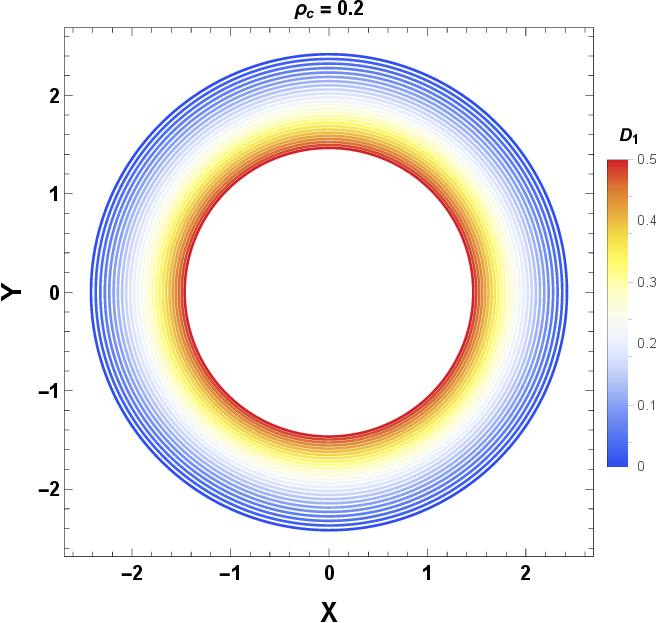, width=.45\linewidth,
height=2.2in}\;\;\; \epsfig{file=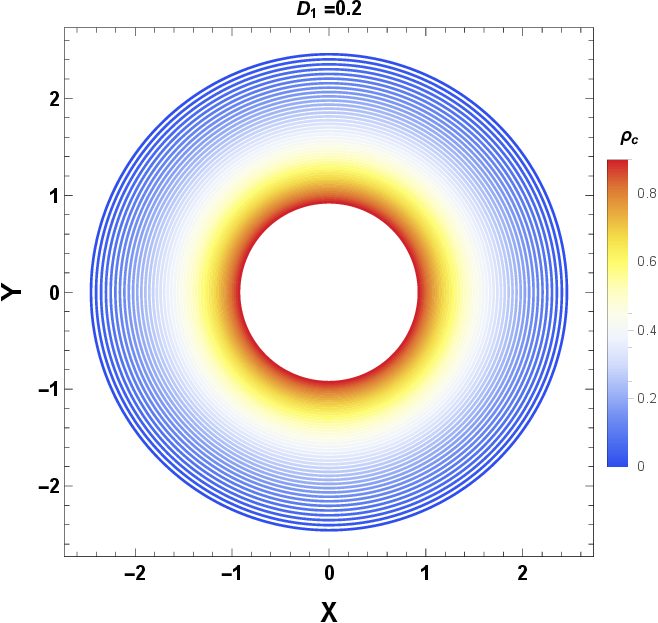, width=.45\linewidth,
height=2.2in}
\centering \epsfig{file=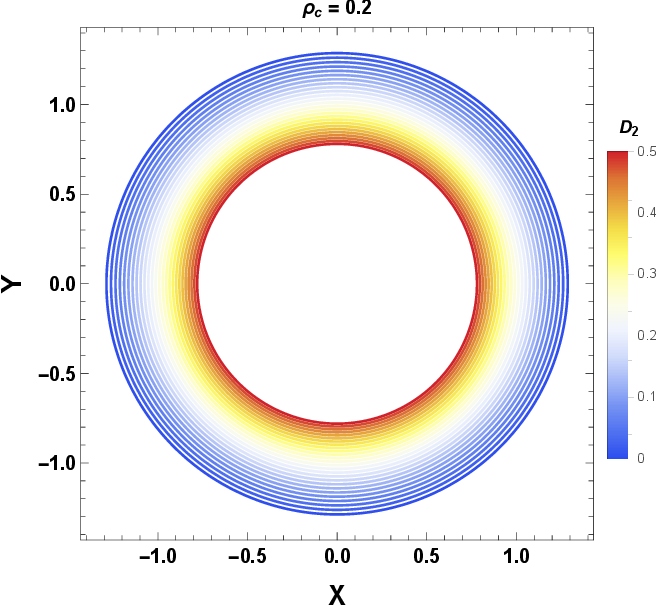, width=.45\linewidth,
height=2.2in}\;\;\; \epsfig{file=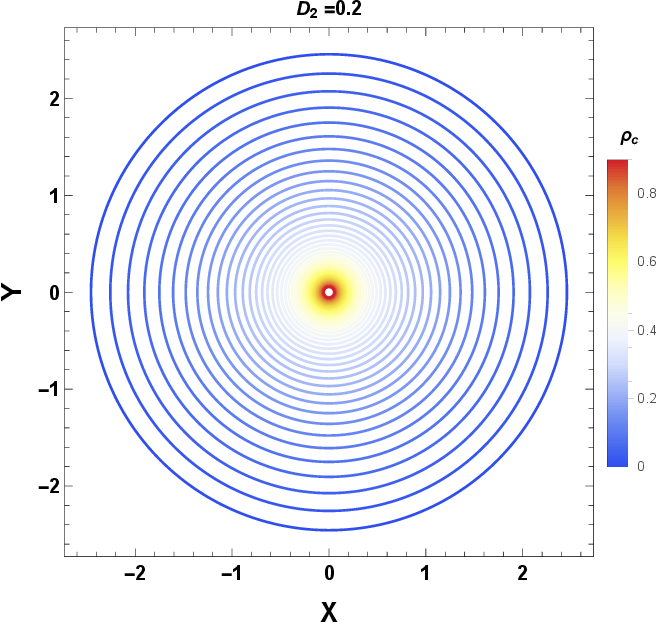, width=.45\linewidth,
height=2.2in}
\centering \epsfig{file=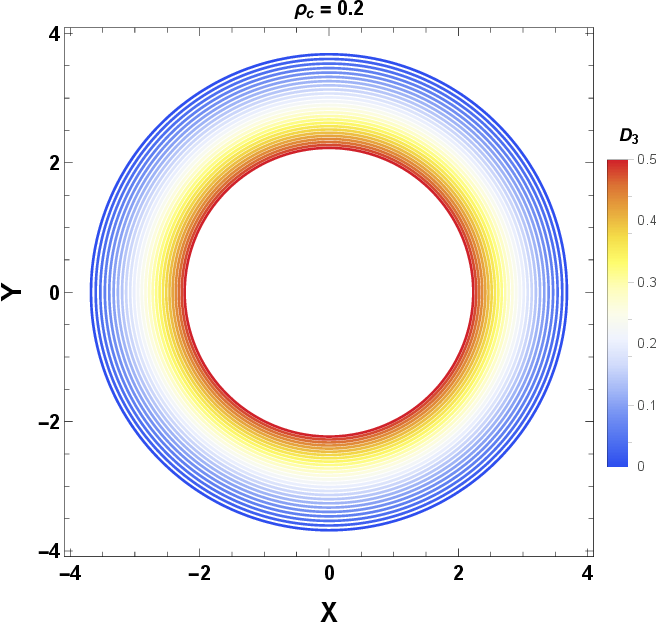, width=.45\linewidth,
height=2.2in}\;\;\; \epsfig{file=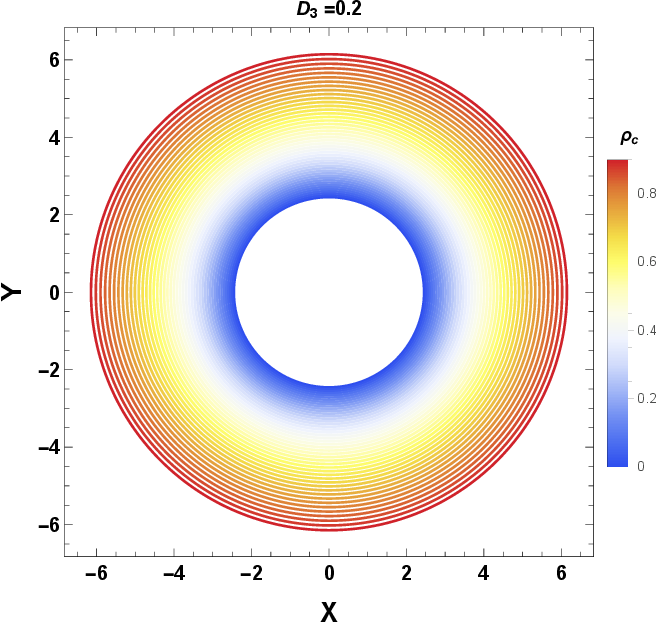, width=.45\linewidth,
height=2.2in}
\caption{\label{rrF8} Picture represents the parametric plots for wormhole shadows with BEC profile (first row), PI profile (second row), and NFW profile (third row) with $r_{0}=0.2$, $R=1.94$, and $r_{s}=1.94$.}
\end{figure*}

In order to describe the wormhole shadows around the wormhole throat, one can use celestial coordinates ($X,Y$), which are further defined as \cite{ws1,ws2,ws3}:
\begin{equation}\label{19}
X=\lim_{r_0\rightarrow \infty}(-r_0^2  \sin(\theta_0))\frac{d\phi}{dr},
\quad
Y=\lim_{r_0\rightarrow \infty}(r_0^2  \frac{d\theta}{dr}),
\end{equation}
where $r_0$ is the wormhole throat. Further, $\theta_0$ is the inclination angle between the wormhole and the observer. After simplification, one can obtain the celestial coordinates for wormhole geometry as \cite{ws4}
\begin{equation}\label{rrr37}
X=-\frac{\xi}{\sin(\theta_0)},
\quad
Y=\sqrt{\eta-\frac{\xi^2}{\sin^2(\theta_0)}}.
\end{equation}
Assuming the static observer is at infinity, the radius of the wormhole shadow $R_s$ as seen from the equatorial plane, i.e., $\theta_0=\pi/2$ can be stated as:
\begin{equation}\label{rrr35}
R_s=\sqrt{\xi^2 +\eta}=\frac{r_0}{e^{\phi(r_0)}}.
\end{equation}
The parametric plot for the equations (\ref{rrr37}) and (\ref{rrr35}) in the $(X, Y)$ plane can cast a variety of wormhole shadows for the different ranges of involved parameters. The wormhole shadows for the wormhole geometry are depicted in Fig. (\ref{rrF8}) for three models of dark matter halos, including BEC. From the first row of Fig. (\ref{rrF8}), it is noticed that the BEC profile has an influence on the wormhole shadows. The larger values of $D_1$ and central density $\rho_c$ lead the wormhole shadow closer to the wormhole throat. For the smaller values of these mentioned parameters wormhole shadow radius is going away from the wormhole throat. The same behavior is also noticed for two other cases, the PI profile and NFW profile from the second row and third row of Fig. (\ref{rrF8}).
\section{Deflection angle}\label{sec6}
\label{secDef}
\begin{figure*}
\centering 
\epsfig{file=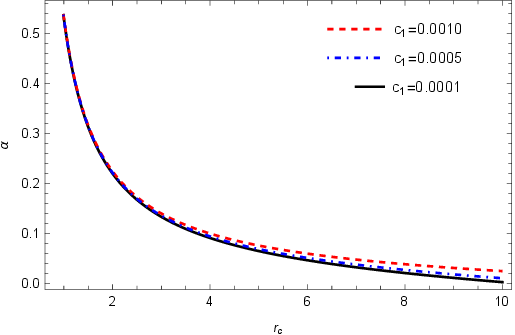, width=.3\linewidth,
height=1.5in}\;\;\; \epsfig{file=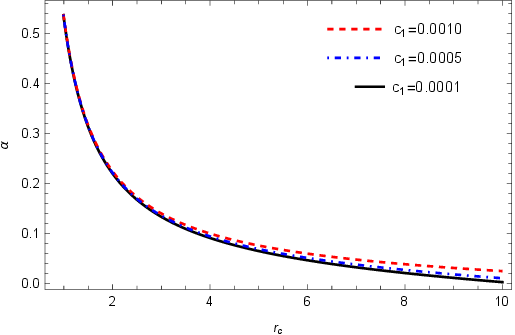, width=.3\linewidth,
height=1.5in}
\centering \epsfig{file=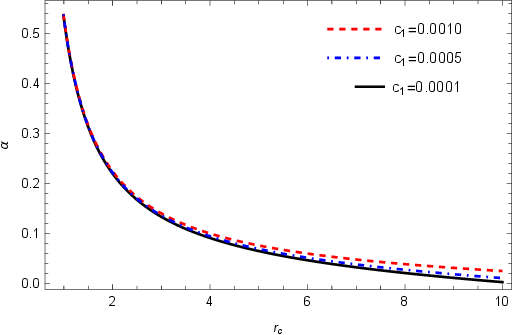, width=.3\linewidth,
height=1.5in}
\caption{\label{rrrF2} Picture represents the deflection angle for BEC (left), PI (middle), and NFW (right) profiles with $c_{0}=0.05$, $R=1.94$, $r_{0}=0.2$, $\rho_s=0.9$, and $r_{s}=1.94$.}
\end{figure*}

This section deals with the deflection angle cast by wormhole geometry under the effect of three different kinds of dark matter halos. The mass and energy generate the curvature of space-time, which influences the speed of light. The curvature of space-time can cause light to bend near a big object, such as a black hole or wormhole. The fascination among researchers with gravitational lensing, especially its strong form, has seen a notable increase following the works by Virbhadra and colleagues \cite{Virbhadra1, Virbhadra3}. Additionally, Bozza, in Ref. \cite{Bozza22}, introduced an analytical approach to compute gravitational lensing in the strong field limit for any spherical symmetric space-time. This method has been applied in numerous subsequent studies, such as Refs. \cite{Quinet, Tejeiro}. This background encourages us to use this analytical technique in our current research. We adopt a numerical technique to study the deflection angle near the wormhole's throat to achieve this aim. For a detailed derivation of the deflection angle, one can refer to the Refs. \cite{g1,r4}.
The formula for deflection angle $\alpha$ for the Morris-Thorne Wormhole geometry can be read as \cite{Bozza1}
\begin{equation}\label{def.angle}
    \alpha=-\pi+2\int_{r_c}^{\infty} \frac{e^{\phi(r)} dr}{r^2 \sqrt{(1-\frac{b(r)}{r})(\frac{1}{\beta^2}-\frac{e^{2\phi(r)}}{r^2})}},
\end{equation}
where $r_s$ is the closest path of light near the throat and $\beta$ is the impact parameter. For null geodesic, we have the relation between $\beta$ and $r_c$, defined by
\begin{equation}
\beta=r_{c} e^{-\phi(r_c)}.    
\end{equation}
We obtained redshift and shape functions in the previous sections for three dark matter models. Here, we shall use those redshift and shape functions to study the deflection angles around the throat.
We substitute the three different sets of redshift functions by Eq. (\ref{39}), Eq. (\ref{4a4}), and Eq. (\ref{53}) and shape functions by Eq. (\ref{41}), Eq. (\ref{48}), and Eq. (\ref{55}) into Eq.(\ref{def.angle}) and solve numerically, one can get deflection angle $\alpha$ with respect to $r_c$ for three different backgrounds. Fig. (\ref{rrrF2}) shows that the deflection angle tends to zero as the distance $r_c$ increases to infinity. In other words, as the light ray goes away from the wormhole throat, where the gravitational field of the wormhole is negligible, it does not deflect from the original path. Conversely, when the value of the distance $r_c$ is close to the radius of the wormhole throat, the deflection angle increases significantly and is positive. Also, in the wormhole throat, where the gravitational field is extremely strong, the deflection of light tends to infinity.

\section{Embedding diagram}\label{sec7}
\begin{figure*}
\centering 
\epsfig{file=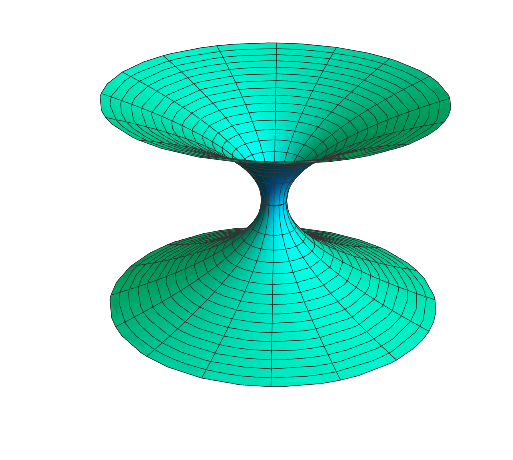, width=.32\linewidth,
height=2in}\;\;\; \epsfig{file=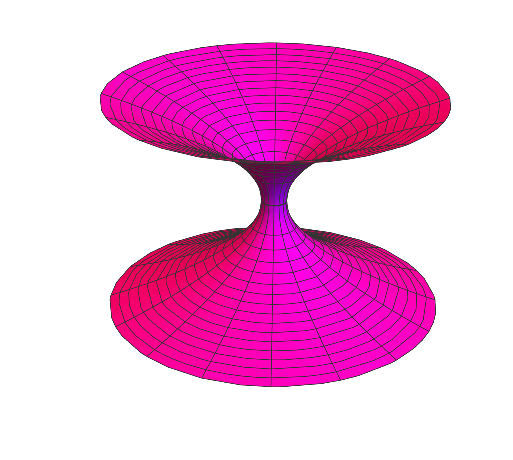, width=.32\linewidth,
height=2in}\;\;\; \epsfig{file=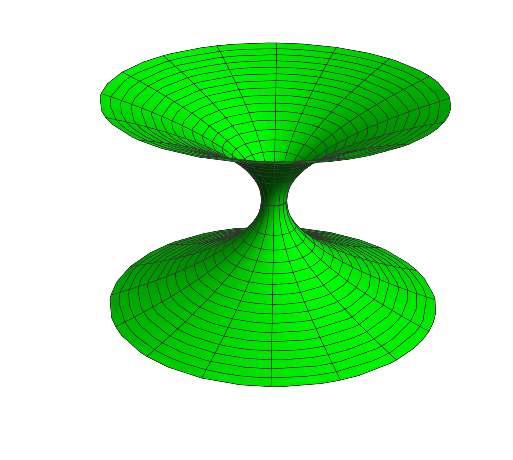, width=.32\linewidth,
height=2in}
\centering 
\epsfig{file=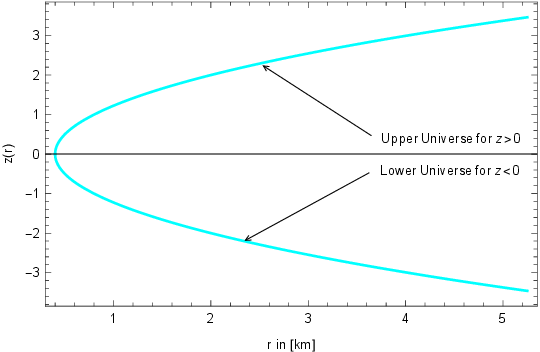, width=.3\linewidth,
height=1.5in}\;\;\; \epsfig{file=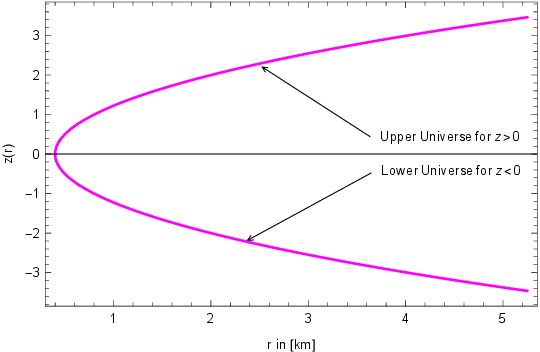, width=.3\linewidth,
height=1.5in} \;\;\; \epsfig{file=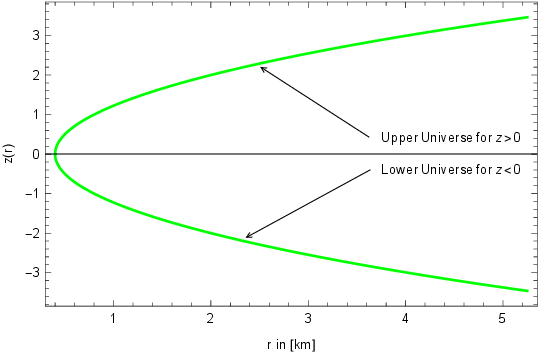, width=.3\linewidth,
height=1.5in} 
\caption{\label{F3} Picture represents the connection of upper and lower Universes through the embedding surface with $c_{0}=0.05$, $R=1.94$, $r_{0}=0.2$, $\rho_s=0.9$, and $r_{s}=1.94$.}
\end{figure*}

\justifying
In this segment, we shall delve into the use of embedding diagrams to gain insights into the structure of wormhole space-time, as referenced in Eq. \eqref{3a}. Our focus is squarely on geometry, which leads us to impose certain constraints on the choice of coordinates. We set $\theta=\pi/2$ on the equatorial plane and fixing time ($t=$ constant). Under these conditions, Eq. \eqref{3a} simplifies to
\begin{equation}
\label{7a1}
ds^2=\left(1-\frac{b(r)}{r}\right)^{-1}+r^2 d\Phi^2.
\end{equation}
We then adapt this modified metric to fit within a three-dimensional Euclidean framework, employing cylindrical coordinates $(r,\, \Phi, \,z)$, which yields
\begin{equation}
\label{7a2}
ds^2=dz^2+dr^2+r^2 d\Phi^2.
\end{equation}
By comparing the two equations above, we deduce the shape of the embedding surface $z(r)$, leading us to establish a gradient as follows:
\begin{equation}
\label{7a3}
\frac{dz}{dr}=\pm \sqrt{\frac{r}{r-b(r)}-1}.
\end{equation}
Eq. \eqref{7a3} shows that the embedded surface becomes vertical at the throat, i.e., $\frac{dz}{dr}$ approaches infinity. Furthermore, it is observed that as r increases towards infinity, indicating the distance from the throat, the curvature, represented by $\frac{dz}{dr}$, tends towards zero, suggesting the space becomes flat. Now, substituting the shape functions given in Eqs. (\ref{41}), (\ref{48}), and (\ref{55}) into the Eq. \eqref{7a3}, we have plotted the embedding diagram, which can be found in Fig. \ref{F3}. In these figures, a positive curvature, $z>0$, denotes the upper universe, while a negative curvature, $z<0$, represents the lower universe.
\section{Conclusions}\label{sec8}
\indent In this study, we have uncovered a novel wormhole solution sustained by DM frameworks such as BEC, PI, and NFW within the $f(Q)$ gravity theory framework. Specifically, our approach involved utilizing the density profile equations of the DM frameworks in conjunction with the rotational velocity to determine both the redshift and shape functions of the wormholes. It is critical to highlight that the model's parameters significantly impact the investigation of the shape of the wormhole. Our findings demonstrate that choosing particular parameters, including the wormhole throat radius $r_0$, leads to a wormhole solution that meets the flare-out condition at the throat, maintaining this characteristic under an asymptotic background. Further, we have investigated the energy conditions using the same parameter involved in the shape functions. Mathematically, we have calculated the expressions of NEC for each model at the throat (see Eq. \eqref{62}). Also, we have depicted the positive and negative regions of all the energy conditions in Figs. (\ref{F5}-\ref{F7}) as well as summarized in Table- \ref{Table1}.\\
\indent Furthermore, we have investigated the wormhole shadow under the effect of DM models. Our findings indicate that the BEC DM model influences the wormhole shadow. Specifically, larger values of $D_1$ and central density $\rho_s$ bring the wormhole shadow closer to the throat, while smaller values of these parameters push the shadow radius away from the throat. We have also observed similar behavior for the PI and NFW profiles.\\
In addition to the shadow, we have extensively analyzed gravitational lensing, specifically the strong gravitational lensing resulting from wormhole geometry under each DM model. The method used in this analysis was developed by Bozza et al. \cite{Bozza1} to explore black hole physics within the strong field regime. Following this, Bozza \cite{Bozza22} advanced an analytical approach to derive gravitational lensing for a general spherically symmetric metric under strong field conditions. This approach has been applied in various studies \cite{r1,r2,r3,r4} to investigate the behavior of deflection angles around wormholes within both GR and modified gravity frameworks.
Consistent with this methodology, we explored the convergence of the deflection angle using the derived solutions for the shape and redshift functions in the context of $f(Q)$ gravity. Our findings reveal that the deflection angle of an outward light ray diverges precisely at the wormhole's throat, representing the surface's correspondence to the photon sphere. Detecting a photon sphere near the wormhole's throat would be highly significant due to its implications. It would confirm the presence of a strong gravitational field and support theoretical predictions related to wormholes. From an observational astronomy standpoint, such a phenomenon would offer a unique opportunity to directly study and observe these mysterious structures, thereby enhancing our understanding of gravity and the fundamental nature of wormholes.\\
In recent decades, numerous researchers have developed various models to explore the presence of wormholes within the framework of Einstein's gravity and modified gravitational theories in the galactic halos. Rahaman and his collaborators, for instance, investigated the potential existence of wormholes within the galactic halo by employing the NFW \cite{Rahaman4} and URC dark matter density profiles \cite{Rahaman2}, utilizing a redshift function derived from a flat rotational curve. Additionally, the Bose-Einstein dark matter density profile has been shown to support the presence of wormholes in the galactic halo \cite{Jusufi1}. Moreover, the geometry of wormholes has been explored with embedded class-I wormholes using URC, NFW, and SFDM dark matter models in Einstein cubic gravity \cite{c4}, as well as in 4D EGB gravity \cite{c5}. In this study, we have employed the BEC, NFW, and PI dark matter models to generate new wormhole solutions within the context of \(f(Q)\) gravity. We have identified more physically viable wormhole solutions by calculating redshift functions derived from flat rotational curves for each dark matter model. Furthermore, we have explored the deflection angles and shadows cast by these wormholes for each dark matter model, marking a novel investigation within the scope of modified gravitational theories.\\
In summary, this study presents a novel wormhole solution within the $f(Q)$ gravity framework, sustained by DM models such as BEC, PI, and NFW. Using the DM density profiles and rotational velocity, we determine the wormhole's redshift and shape functions, highlighting the significant influence of model parameters on the wormhole's shape, particularly the throat radius $r_0$. The energy conditions are evaluated at the throat, revealing regions where these conditions are met or violated. The study also explores the wormhole shadow, showing that larger values of specific parameters bring the shadow closer to the throat in the BEC DM model, a trend similarly observed in PI and NFW profiles. Additionally, gravitational lensing effects are examined, demonstrating that the deflection angle increases near the wormhole throat, where the gravitational field is strongest, and tends to zero as the distance from the throat increases, indicating minimal light deviation in weaker gravitational fields.\\
\indent Thus, it is safe to conclude that our findings suggest the possibility of the existence of wormholes in the galactic halos caused by BEC, PI, and NFW dark matter within the framework of $f(Q)$ gravity. Additionally, the effect of these dark matter models could be explored in other modified gravity theories, such as $f(T)$ gravity, as both $f(T)$ and $f(Q)$ models are indistinguishable at the cosmological background level \cite{Lin2}.

\section*{Data Availability}
There are no new data associated with this article.

\acknowledgments ZH acknowledges the Department of Science and Technology (DST), Government of India, New Delhi, for awarding a Senior Research Fellowship (File No. DST/INSPIRE Fellowship/2019/IF190911). PKS acknowledges the National Board for Higher Mathematics (NBHM) under the Department of Atomic Energy (DAE), Govt. of India, for financial support to carry out the Research project No.: 02011/3/2022 NBHM(R.P.)/R\&D II/2152 Dt.14.02.2022.

\end{document}